\documentclass[twocolumn,english,nofootinbib,showpacs,superscriptaddress,pra,preprintnumbers,amsmath,amssymb,floatfix,longbibliography]{revtex4-2}
\usepackage[T1]{fontenc}
\usepackage[utf8]{inputenc}
\usepackage{color}
\usepackage{babel}
\usepackage{bm}
\usepackage{amsmath}
\usepackage{graphicx}
\usepackage{esint}
\usepackage[unicode=true,pdfusetitle,
 bookmarks=true,bookmarksnumbered=false,bookmarksopen=false,
 breaklinks=false,pdfborder={0 0 0},pdfborderstyle={},backref=false,colorlinks=true]
 {hyperref}

\makeatletter

\usepackage{dcolumn}
\usepackage{bm}
\usepackage{color}
\usepackage{soul}
\usepackage{babel}
\usepackage{amsfonts}
\usepackage{slashed}
\usepackage{enumerate}

\usepackage{babel}

\makeatother

\begin{document}
\global\long\def\sgn{\mathrm{sgn}}%
\global\long\def\ket#1{\left|#1\right\rangle }%
\global\long\def\bra#1{\left\langle #1\right|}%
\global\long\def\sp#1#2{\langle#1|#2\rangle}%
\global\long\def\abs#1{\left|#1\right|}%
\global\long\def\avg#1{\langle#1\rangle}%

\title{Weak-measurement-induced phases and dephasing: broken symmetry of
the geometric phase}
\author{Kyrylo Snizhko}
\affiliation{Department of Condensed Matter Physics, Weizmann Institute of Science,
Rehovot, 76100 Israel}
\affiliation{Institute for Quantum Materials and Technologies, Karlsruhe Institute
of Technology, 76021 Karlsruhe, Germany}
\author{Nihal Rao}
\affiliation{Department of Condensed Matter Physics, Weizmann Institute of Science,
Rehovot, 76100 Israel}
\affiliation{Present affiliation: Arnold Sommerfeld Center for Theoretical Physics,
University of Munich, Theresienstr. 37, 80333 M\"unchen, Germany}
\affiliation{Present affiliation: Munich Center for Quantum Science and Technology
(MCQST), Schellingstr. 4, 80799 M\"unchen, Germany}
\author{Parveen Kumar}
\affiliation{Department of Condensed Matter Physics, Weizmann Institute of Science,
Rehovot, 76100 Israel}
\author{Yuval Gefen}
\affiliation{Department of Condensed Matter Physics, Weizmann Institute of Science,
Rehovot, 76100 Israel}
\begin{abstract}
Coherent steering of a quantum state, induced by a sequence of weak
measurements, has become an active area of theoretical and experimental
study. For a closed steered trajectory, the underlying phase factors
involve both geometrical and dynamical terms. Furthermore, considering
the reversal of the order of the measurement sequence, such a phase
comprises a symmetric and an antisymmetric term. Superseding common
wisdom, we show that the symmetric and the antisymmetric components
do not correspond to the dynamical and geometrical parts respectively.
Addressing a broad class of measurement protocols, we further investigate
the dependence of the induced phases on the measurement parameters
(e.g., the measurement strength). We find transitions between different
topologically distinct sectors, defined by integer-valued winding
numbers, and show that the transitions are accompanied by diverging
dephasing. We propose experimental protocols to observe these effects.

\emph{}
\end{abstract}
\maketitle

\section{Introduction.}

Geometrical phases are a cornerstone of modern physics \citep{Cohen2019}.
The work of Berry \citep{Berry1984} provided a unifying language
that is key to understanding disparate phenomena including the quantum
Hall effect \citep{Thouless1982,Niu1985} and topological insulators
\citep{Hasan2010}, sheds light on some features of graphene \citep{Zhang2005,Novoselov2006},
and provides the basis for geometric \citep{Pachos1999,Jones2000}
and topological \citep{Nayak2008} quantum computation platforms.
Geometrical phases can be induced not only by means of adiabatic \citep{Berry1984}
or non-adiabatic \cite{Aharonov1987, [See also ] Wood2020} Hamiltonian
manipulation, but also as a result of a sequence of projective measurements
\citep{Berry1996,Facchi1999a,Chruscinski2004}. In that case, the
phase is called the Pancharatnam phase, after the Indian physicist
who discovered it in the context of classical optics \citep{Pancharatnam1956}.
Recently, the possibility of inducing geometrical phases by weak measurements
\citep{Tamir2013,Svensson2013} was demonstrated experimentally \citep{Cho2019}.
Moreover, a topological transition in the behavior of the geometrical
phase as a function of the measurement strength has been predicted
theoretically \citep{Gebhart2020}.

Here we outline a general framework for treating measurement-induced
phase factors and\textcolor{blue}{{} }apply it to a broad class of measurements.
Our analysis addresses the nature of the phase accumulated during
a sequence of weak measurements, a generalization of the concept of
the geometrical Pancharatnam phase in the case of strong (projective)
measurements. In previous investigations, measurement-induced phase
factors were of a purely geometric origin \citep{Berry1996,Cho2019,Gebhart2020}.
In the presence of an additional Hamiltonian acting on the measured
system, an additional dynamical component appears \citep{Facchi1999a}.
We demonstrate that weak-measurement-induced phases generically involve
both geometrical and dynamical components even in the absence of an
additional Hamiltonian. This fact went unnoticed in earlier studies
that focused on restricted classes of measurements. Quantum measurements
are characterized by Kraus operators describing the consequent back-action
\citep{Nielsen2010,Wiseman2010,Jacobs2014a}. While previous works
focused on the case of Hermitian Kraus operators, in the more general
case of non-Hermitian Kraus operators, considered here, not all of
measurement-induced phases can qualify as geometrical.

We also investigate the behavior of the phase with respect to reversing
the order of the measurement sequence. The fact that there is no definite
symmetry with respect to such a reversal implies that measurement-induced
phases can be split into a symmetric and an antisymmetric components.
Interestingly, and superseding the common structure of phases generated
by conventional adiabatic Hamiltonian dynamics, the symmetric and
antisymmetric components do not coincide with the dynamical and geometrical
components respectively.

These general insights are then demonstrated through the analysis
of specific measurement protocols. We study two types of such protocols:
one which involves postselection, and a second which involves averaging
over all measurement outcomes (i.e., no postselection). Postselection
refers to selecting experimental runs that yield a desired set of
measurement readouts. An important quantity here is the postselection
probability, i.e., the probability to have a predesignated readout
sequence. Concerning the other protocol, one averages the readout-sequence-dependent
phase over many experimental runs, which gives rise to a suppression
factor a.k.a. dephasing.

Finally, we focus on topological transitions in the context of measurement-induced
phases. Previously, such transitions were predicted for a restricted
class of measurements with Hermitian back-action \citep{Gebhart2020}.
Here we show that such transitions may still take place when the backaction
is non-Hermitian. Under such general back-action, multiple distinct
topological sectors exist, forming a rich ``phase diagram''. Transitions
between such sectors are marked by (i) a vanishing probability of
the corresponding postselected sequence (the case of postselective
protocols); (ii) diverging dephasing (the case of averaging protocols).\footnote{A short exposition of our results pertaining to topological transitions
in the averaging protocols can be found in Ref.~\citep{Snizhko2020b}.} We also propose and analyze experimental setups that may test our
predictions.

The paper is organized as follows. Section~\ref{sec:II_definitions_analysis}
first recaps the theory of generalized quantum measurements. We then
define measurement-induced phases, and discuss their classifications
into dynamical/geometrical and symmetric/antisymmetric terms and the
relation between these two classifications. In Section~\ref{sec:III_particular_protocol}
we specify the measurements and protocols to be employed. We derive
and analyze analytic expressions for the induced phases, the postselection
probabilities, and the dephasing factors. Section~\ref{sec:IV_top_trans_postselected}
presents a mostly numerical analysis of the topological transitions
vis-a-vis postselective protocols. Section~\ref{sec:V_top_trans_averaged}
presents a similar analysis for the phase-averaging protocol. In Section~\ref{sec:VI_experimental_implementation}
we discuss possible experimental implementations. Conclusions are
presented in Section~\ref{sec:VII_Conclusions}. Three appendices
of technical nature are included. Appendix~\ref{sec:appendix_scaling_regimes}
presents a justification of our choice of scaling of the measurement
parameters with the number of measurements, cf.~Sec.~\ref{sec:III_particular_protocol}.
Appendix~\ref{sec:appendix_critical_line_postselected} provides
an analytic derivation of the critical line of topological transitions
in the postselective protocol, cf.~Sec.~\ref{sec:IV_top_trans_postselected}.
Appendix~\ref{sec:appendix_averaged_phase_detection} provides the
justification for the averaged phase detection scheme proposed in
Sec.~\ref{sec:VI_experimental_implementation}.

\section{\label{sec:II_definitions_analysis}Weak-measurement-induced phases:
definitions and general analysis}

In this section, we present a general analysis of measurement-induced
phase factors. We briefly recall the theory of generalized quantum
measurements in Sec.~\ref{sec:IIA_measurement_theory}. We then proceed
to define postselected and averaged measurement-induced phases in
Sec.~\ref{sec:IIB_phases_definitions}. We analyze various characteristics
of these phases and discuss possible classifications thereof in Sec.~\ref{sec:IIC_dyn/geom+sym/antisym_classifications}.

\subsection{\label{sec:IIA_measurement_theory}Theory of generalized measurements}

Describing a conventional projective measurement in quantum theory
requires a Hermitian observable $\mathcal{O}$ of the measured system.
The observable has a set of eigenstates labeled by its eigenvalues
$\lambda$, $\mathcal{O}\ket{\lambda}=\lambda\ket{\lambda}$. A projective
measurement yields a readout $r$ which corresponds to one of the
eigenvalues $\lambda$. If a readout $r=\lambda$ is obtained, the
system state becomes $\ket{\psi^{(r=\lambda)}}=\mathcal{P}_{\lambda}\ket{\psi}$,
where $\ket{\psi}$ is the system state before the measurement and
$\mathcal{P}_{\lambda}=\ket{\lambda}\bra{\lambda}$ is the projector
onto the corresponding eigenstate of $\mathcal{O}$ (generalization
to the case of a degenerate spectrum is straightforward). Note that
$\ket{\psi^{(r=\lambda)}}$ is not normalized. The probability of
the projective measurement yielding $r=\lambda$ is $p_{r=\lambda}=\abs{\sp{\lambda}{\psi}}^{2}=\sp{\psi^{(r=\lambda)}}{\psi^{(r=\lambda)}}$.

Generalized measurement \citep{Nielsen2010,Wiseman2010,Jacobs2014a}
is an extension of the orthodox concept of projective measurement.
The extension is based on treating the detector as an additional quantum-mechanical
object. The measurement is then conceptually described as a two-step
protocol: (i) the system is coupled to the detector, and then decoupled;
(ii) the detector is measured projectively. The strength of the interaction
between the system and the detector defines the measurement strength.

The formal description of such a protocol is as follows: Let the system
initial state be $\ket{\psi}$ in the system Hilbert space $\mathcal{H}_{\mathrm{s}}$
and the detector initial state be $\ket{D_{i}}$ in the detector Hilbert
space $\mathcal{H}_{\mathrm{d}}$. During the first step, they interact
via Hamiltonian $H_{\mathrm{s-d}}(t)$ which vanishes outside the
interval $t\in[0;T]$ (i.e., the interaction Hamiltonian is switched
on at $t=0$ and off at $t=T$). In the second step, the detector
is measured projectively with readouts corresponding to some basis
$\left\{ \ket r\right\} $ in the detector's Hilbert space. The outcome
of the first step is the evolution of the system-detector state
\begin{multline}
\ket{\psi}\ket{D_{i}}\rightarrow\\
\mathcal{T}\exp\left(-i\int_{0}^{T}H_{\mathrm{s-d}}(t)dt\right)\ket{\psi}\ket{D_{i}}=\sum_{r}\ket{\psi^{(r)}}\ket r,
\end{multline}
where $\mathcal{T}$ stands for time ordering; the last equality represents
a decomposition that can be performed for any pure state in the system-detector
Hilbert space $\mathcal{H}_{\mathrm{s}}\otimes\mathcal{H}_{\mathrm{d}}$.
For a specific system-detector Hamiltonian $H_{\mathrm{s-d}}(t)$
and detector's initial state $\ket{D_{i}}$, the resulting system
state can be written as\textcolor{blue}{{} }\citep{Wiseman2010}
\begin{equation}
\ket{\psi^{(r)}}=\mathcal{M}^{(r)}\ket{\psi},\label{eq:gen_measurement_back_action}
\end{equation}
where the Kraus operators 
\begin{equation}
\mathcal{M}^{(r)}=\bra r\mathcal{T}\exp\left(-i\int_{0}^{T}H_{\mathrm{s-d}}(t)dt\right)\ket{D_{i}}\label{eq:Kraus_microscopic_expression}
\end{equation}
represent the measurement's non-local backaction, following the detector's
projective readout. The probability of obtaining a specific readout
$r$ is
\begin{equation}
p_{r}=\sp{\psi^{(r)}}{\psi^{(r)}}=\bra{\psi}\mathcal{M}^{(r)\dagger}\mathcal{M}^{(r)}\ket{\psi}.\label{eq:readout_probability_general}
\end{equation}
Conservation of probability, $\sum_{r}p_{r}=1$, independently of
the system's initial state, $\ket{\psi}$, implies
\begin{equation}
\sum_{r}\mathcal{M}^{(r)\dagger}\mathcal{M}^{(r)}=\mathbb{I},
\end{equation}
where $\mathbb{I}$ is the identity operator acting in the system's
Hilbert space. This is the only restriction on the Kraus operators,
which otherwise are arbitrary.\footnote{Anticipating the use of multiple measurements in our protocols below,
we comment on the relation between the dynamics of a system under
measurement and the dynamics of an open system. Indeed, a detector
can be regarded as the system's environment, putting the measurement-induced
dynamics into the framework of open quantum systems. However, the
formal analogy is incomplete without emphasizing a key difference
between the meaurement-induced and open-system dynamics. The standard
treatement of an open system does not consider modulating the system-environment
coupling, nor directly controlling the environment in the course of
experiment. As a result, a classification into Markovian and non-Markovian
open system dynamics (depending on the environment properties) arises.
In contrast, measurement-based dynamics includes the following aspects:
modulating system-detector couplings to separate distinct measurements,
reading out the detector, and preparing the detector in a specific
initial state before the next measurement. As a result, the measurement-induced
dynamics is non-Markovian on the scale of a single measurement (as
the detector does not relax and thus preserves memory of the system's
previous states during the measurement), but is Markovian beyond the
single-measurement timescale (the detector is read out and initialized
before each measurement; therefore, the result of a measurement depends
only on the system state at the beginning of the measurement, and
not on the previous system states/measurement readouts).}

One thus sees that a description of a generalized quantum measurement
does not require microscopic modeling of the detector. It is sufficient
to specify the set of possible readouts $\left\{ r\right\} $ and
the corresponding Kraus operators $\mathcal{M}^{(r)}$ acting on the
system. The Kraus operators are thus the analogues of the projection
operators $\mathcal{P}_{\lambda}$ that describe the back-action of
a projective measurement.\textcolor{blue}{{} }If the Kraus operators
$\mathcal{M}^{(r)}$ are Hermitian and $(\mathcal{M}^{(r)})^{2}=\mathcal{M}^{(r)}$
then they can be interpreted as projectors $\mathcal{P}_{r}$ and
the generalized measurement scheme reduces to the projective measurement
scenario.

\begin{figure}
\begin{centering}
\includegraphics[width=0.6\columnwidth]{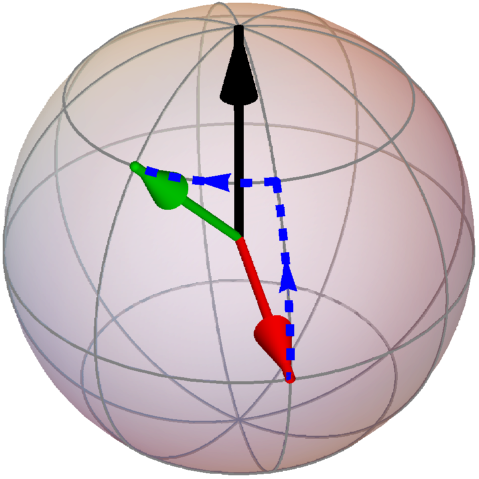}
\par\end{centering}
\caption{\label{fig:measurement_back_action}Back-action of a generalized measurement.
After a projective measurement of the $S_{z}$ component of spin $1/2$
that yields a readout $S_{z}=+1/2$, the initial state (red arrow)
becomes aligned with the north pole of the Bloch sphere (black arrow).
A generalized measurement's back-action does not necessarily align
the state with an eigenstate of the measured observable; it only pulls
the state towards the north pole and may also rotate it around the
$z$ axis (green arrow). These two effects of the back-action are
illustrated by blue dashed lines. Such a back-action appears in the
measurement protocol we consider in Sec.~\ref{sec:IIIA_detector_model}.}
\end{figure}

It is important to realize the following difference between projective
and generalized measurements. In a projective measurement, a specific
readout $r=\lambda$ always corresponds to the system being collapsed
onto a specific state $\ket{\lambda}\propto\mathcal{P}_{\lambda}\ket{\psi}$
following the measurement. This is not necessarily so for generalized
measurements. Indeed, knowing the initial state $\ket{\psi}$ and
the readout $r$, one finds the system state after the measurement
according to Eq.~(\ref{eq:gen_measurement_back_action}). However,
knowing the measurement readout alone does not suffice for the determination
of $\ket{\psi^{(r)}}$. An example of the generalized measurement
back-action is presented in Fig.~\ref{fig:measurement_back_action}.

It is instructive to understand how the system observable measured,
the system-detector interaction Hamiltonian, and the Kraus operators
are related. For that, consider a simple canonical example of the
system-detector Hamiltonian, 
\begin{equation}
H_{\mathrm{s-d}}(t)=\lambda\,\theta_{[0;T]}(t)\;\mathcal{O}\otimes A^{(\mathrm{d})},
\end{equation}
where $\mathcal{O}$ is the system observable measured, $A^{(\mathrm{d})}$
is an operator acting in the detector Hilbert space $\mathcal{H}_{\mathrm{d}}$,
$\lambda$ is the coupling strength, and
\begin{equation}
\theta_{[0;T]}(t)=\begin{cases}
0, & \text{if }t<0\text{ or }t>T,\\
1, & \text{if }0<t<T.
\end{cases}
\end{equation}
Note that the same measurement setup (the same system and detector,
same detector initial state, same detector readout basis) can be used
with different observables $\mathcal{O}$. In particular, with observables
of the form $R^{-1}\mathcal{O}R$, where $R$ is a unitary rotation
in the system Hilbert space. The back-action of the measurement then
changes 
\begin{equation}
\mathcal{M}^{(r)}\rightarrow R^{-1}\mathcal{M}^{(r)}R.\label{eq:same-type_Kraus}
\end{equation}
Equation~(\ref{eq:same-type_Kraus}) defines a family of measurements
of the ``same class''. Modifying the system-detector interaction,
e.g., by selecting an observable $\mathcal{O}'$ with a different
spectrum, modifies the nature of the measurement at hand, thus introducing
a different measurement class. The measurement class may be altered
even more drastically, e.g., by keeping the same observable but taking
a different detector (with different operator $A^{(\mathrm{d})}$,
different initial state $\ket{D_{i}}$, different readout basis $\left\{ \ket r\right\} $
or even different Hilbert space $\mathcal{H}_{\mathrm{d}}$). Thus,
measurements of different classes can apply to the same observable
$\mathcal{O}$ but yield drastically different back-actions or even
have different sets of possible readouts $r$. In principle, nothing
prevents us from implementing measurements of different classes on
a given system at different times.

\subsection{\label{sec:IIB_phases_definitions}Measurement-induced phases}

Consider a sequence of $N+1$ distinct measurements performed on a
quantum system. Each measurement is fully characterized by a set of
Kraus operators, $\left\{ \mathcal{M}_{k}^{(r_{k})}\right\} $, where
$k=1,...,N+1$ is the measurement number and $r_{k}$ is the measurement
readout. These can be measurements of the same class, yet measuring
different system observables (e.g., same strength measurements of
the spin projection onto different directions), in which case
\begin{equation}
\mathcal{M}_{k}^{(r_{k})}=R_{k}^{-1}M^{(r_{k})}R_{k},
\end{equation}
where $R_{k}$ is a unitary rotation in the system Hilbert space and
the Kraus operator $M^{(r)}$ does not depend on $k$. This is the
case for the example considered in Ref.~\citep{Gebhart2020} and
for the one discussed in Sec.~\ref{sec:III_particular_protocol}.
However, in the present section we keep the analysis general. In particular,
we allow for situations where the Kraus operators for different measurements
are not simply related and even the number of possible readout values,
$r_{k}$, for different measurements can be different.

Consider a system prepared in a certain initial state $\ket{\psi_{0}}$.
Assuming knowledge of all readouts $\left\{ r_{k}\right\} $ of the
measurement sequence, the system state traverses a sequence of states
\begin{equation}
\ket{\psi_{k}}=\mathcal{M}_{k}^{(r_{k})}...\mathcal{M}_{2}^{(r_{2})}\mathcal{M}_{1}^{(r_{1})}\ket{\psi_{0}}.\label{eq:def_intermediate_states}
\end{equation}
We choose the last measurement to be projective and postselect the
final state such that it coincides with the system's initial state,
i.e., $\mathcal{M}_{N+1}^{(0)}=\mathcal{P}_{0}=\ket{\psi_{0}}\bra{\psi_{0}}$\textcolor{blue}{.}
Then the system state after completing the entire sequence of measurements,
\begin{equation}
\ket{\psi_{N+1}}=\ket{\psi_{0}}\bra{\psi_{0}}\mathcal{M}_{N}^{(r_{N})}...\mathcal{M}_{1}^{(r_{1})}\ket{\psi_{0}},
\end{equation}
differs from the initial state by a factor
\begin{equation}
\bra{\psi_{0}}\mathcal{M}_{N}^{(r_{N})}...\mathcal{M}_{1}^{(r_{1})}\ket{\psi_{0}}=\sqrt{P_{\{r_{k}\}}}e^{i\chi_{\{r_{k}\}}}.\label{eq:chi_definition}
\end{equation}
This factor has two components: $P_{\{r_{k}\}}$ is the probability
to observe the readout sequence $\left\{ r_{1};...;r_{N};r_{N+1}=0\right\} $,
while $e^{i\chi_{\{r_{k}\}}}$ is the phase factor accrued by the
state due to undergoing measurement-induced evolution. In what follows
we refer to $\chi_{\{r_{k}\}}$ as the postselected measurement-induced
phase.\footnote{It is known that the phase of an individual quantum state can be chosen
arbitrarily, which is often referred to as the gauge freedom. The
phase defined here is gauge-invariant. In order to understand this,
consider the two potential sources of non-gauge-invariance. One source
is the freedom of choosing the phase of the initial state $\ket{\psi_{0}}\rightarrow e^{i\phi}\ket{\psi_{0}}$.
One immediately sees that this freedom does not affect Eq.~(\ref{eq:chi_definition});
the physical reason is that, just as in Ref.~\citep{Berry1984},
the phase is defined via comparison of the state after some process
with the system's initial state. The second source is the gauge freedom
associated with choosing the phase of the detector states $\ket{D_{i}}$
and $\ket r$, which would affect the phase of the the Kraus operators
$\mathcal{M}^{(r)}$, cf.~Eq.~(\ref{eq:Kraus_microscopic_expression}).
This freedom is, however, eliminated by demanding that when the system
does not interact with detectors ($H_{\mathrm{s-d}}(t)\equiv0$),
no phase is accumulated. In the practice (cf.~the experimental setups
proposed in Sec.~\ref{subsec:VI.A_interferometric_detection_schemes}),
both of these theoretical sources of gauge non-invariance are made
inconsequential by the structure of the interferometer, which ensures
that (i) the unaffected initial state is compared to the state after
a sequence of measurements, directly implementing the first theoretical
argument; (ii) the detector preparation and readout is performed independently
of the path taken by the particle, therefore, any phases associated
with $\sp r{D_{i}}$ ``cancel out'' in the interference pattern.} In the case where all the measurements are projective, $\chi_{\{r_{k}\}}$
reduces to the projective-measurement-induced Pancharatnam phase \citep{Pancharatnam1956,Berry1996,Chruscinski2004}.

It is also possible to define the averaged measurement-induced phase
through
\begin{multline}
e^{2i\bar{\chi}-\alpha}=\sum_{\{r_{k}\}}\left(\bra{\psi_{0}}\mathcal{M}_{N}^{(r_{N})}...\mathcal{M}_{1}^{(r_{1})}\ket{\psi_{0}}\right)^{2}\\
=\sum_{\{r_{k}\}}P_{\{r_{k}\}}e^{2i\chi_{\{r_{k}\}}},\label{eq:chi-bar_definition}
\end{multline}
where the sum runs over all possible readout sequences $\left\{ r_{k}\right\} $
(such that $r_{N+1}=0$). $\bar{\chi}$ is the averaged measurement-induced
phase. The real parameter $\alpha\geq0$ has a mixed meaning. It characterizes
the dephasing due to averaging over various measurement readout sequences
$\{r_{k\leq N}\}$; at the same time, the finite probability of obtaining
$r_{N+1}=0$ in the last projective measurement also contributes to
$\alpha$. Hereafter we will refer to $\alpha$ as the dephasing parameter
and to $e^{-\alpha}$ as the dephasing-induced suppression factor.

One may wonder why the averaged phase is defined through the averaging
of $e^{2i\chi_{\{r_{k}\}}}$ in Eq.~(\ref{eq:chi-bar_definition}),
and not through $\sum_{\{r_{k}\}}P_{\{r_{k}\}}e^{i\chi_{\{r_{k}\}}}$.
The reason is rooted in the phase measurement procedure, discussed
in detail in Sec.~\ref{sec:VI_experimental_implementation}. Here
we only briefly explain the idea behind the procedure of observing
the averaged phase. Different readout sequences $\{r_{k}\}$ correspond
to mutually orthogonal states of detectors employed throughout the
sequence of measurements. At the same time, measuring a phase requires
interference between two states, e.g., the unmeasured and the measured
states. For measuring the phase corresponding to a postselected sequence
$\{r_{k}\}$, one can use an interferometer, in one arm of which the
system (spin of the flying particle) is measured, and in the other
it is not, cf.~Fig.~\ref{fig:interferometers}(a). If the initial
state of all detectors coincides with the state corresponding to the
postselected readout sequence, the interference pattern exhibits a
non-vanishing visibility, which allows for measuring $\chi_{\{r_{k}\}}$.
This may work for one particular postselected readout sequence. However,
averaging requires the consideration of numerous readout sequences,
the vast majority of which are orthogonal to the sequence of null
readouts, expected when no system-detector coupling is present (i.e.,
when the interfering particle goes through the reference interferometer
arm which does not involve coupling to detectors). To facilitate averaging
over different readout sequences, one needs to couple detectors to
both arms of the interferometer, cf.~Fig.~\ref{fig:interferometers}(b).
This facilitates maintaining coherence between the two arms independently
of the measurement readouts. In other words: readouts do not constitute
a ``which path'' measurement \citep{Buks1998,Goldstein2016}. We
design the couplings such that traversing one arm of the inteferometer
or the other, the system accumulates opposite phases, $e^{i\chi_{\{r_{k}\}}}$
and $e^{-i\chi_{\{r_{k}\}}}$. As a result, the phase factor that
should be averaged is $e^{2i\chi_{\{r_{k}\}}}$.

\subsection{\label{sec:IIC_dyn/geom+sym/antisym_classifications}Classification
of measurement-induced phases}

In Sec.~\ref{sec:IIB_phases_definitions}, we defined postselected
(\ref{eq:chi_definition}) and averaged (\ref{eq:chi-bar_definition})
measurement-induced phases. Here we investigate their separation into
dynamical and geometrical components and their symmetry properties
with respect to reversing the order of the measurement sequence.

Consider a specific readout sequence, $\left\{ r_{k}\right\} $. The
system traverses a trajectory in the Hilbert space corresponding to
states $\ket{\psi_{k}}$, defined in Eq.~(\ref{eq:def_intermediate_states}).
It is known that any quantum system traversing a trajectory in the
Hilbert space accumulates a Pancharatnam phase \citep{Pancharatnam1956,Berry1996,Chruscinski2004},
\begin{equation}
\arg\bra{\psi_{0}}\mathcal{P}_{N}...\mathcal{P}_{1}\ket{\psi_{0}},\label{eq:Pancharatnam_phase}
\end{equation}
where $\arg$ denotes the argument of a complex number and $\mathcal{P}_{k}=\ket{\psi_{k}}\bra{\psi_{k}}/\sp{\psi_{k}}{\psi_{k}}$
are the projectors onto the respective intermediate states. Does Eq.~(\ref{eq:Pancharatnam_phase})
coincide with $\chi_{\{r_{k}\}}$? In general, Pancharatnam's geometrical
phase (\ref{eq:Pancharatnam_phase}) does not coincide with $\chi_{\{r_{k}\}}$
(\ref{eq:chi_definition}), implying that the latter has a geometrical
and a non-geometric (a.k.a. dynamical) components.

One can articulate a simple condition for $\chi_{\{r_{k}\}}$ to be
purely geometrical. If the Kraus operators used in the measurement
sequence are Hermitian ($\mathcal{M}_{k}^{(r_{k})\dagger}=\mathcal{M}_{k}^{(r_{k})}$)
and positive-semidefinite ($\bra{\psi}\mathcal{M}_{k}^{(r_{k})}\ket{\psi}\geq0$
for any $\ket{\psi}$), then $\sp{\psi_{k+1}}{\psi_{k}}=\bra{\psi_{k}}\mathcal{M}_{k+1}^{(r_{k+1})\dagger}\ket{\psi_{k}}=\bra{\psi_{k}}\mathcal{M}_{k+1}^{(r_{k+1})}\ket{\psi_{k}}\geq0$.
This implies that 
\begin{multline}
\arg\bra{\psi_{0}}\mathcal{M}_{N}^{(r_{N})}...\mathcal{M}_{1}^{(r_{1})}\ket{\psi_{0}}=\arg\sp{\psi_{0}}{\psi_{N}}\\
=\arg\sp{\psi_{0}}{\psi_{N}}+\arg\prod_{k=0}^{N-1}\sp{\psi_{k+1}}{\psi_{k}}\\
=\arg\bra{\psi_{0}}\mathcal{P}_{N}...\mathcal{P}_{1}\ket{\psi_{0}},\label{eq:phase_through_projectors}
\end{multline}
so that $\chi_{\{r_{k}\}}$ coincides with the Pancharatnam phase.\footnote{Note that this Pancharatnam phase is determined by the system's intermediate
states and not by the measurement directions. The latter coincide
with the former only for projective measurements.} Generically, however, the Kraus operators are not Hermitian. Then
$\sp{\psi_{k+1}}{\psi_{k}}=\bra{\psi_{k}}\mathcal{M}_{k+1}^{(r_{k+1})\dagger}\ket{\psi_{k}}$
is not constrained to be real (not to mention non-negative), Eq.~(\ref{eq:phase_through_projectors})
does not hold, and the phases $\chi_{\{r_{k}\}}$ are not uniquely
determined by the measurement-induced state trajectory (although knowledge
of the trajectory together with the measurement parameters clearly
does determine the phase). We call the difference 
\begin{equation}
\chi_{\{r_{k}\}}^{(\mathrm{dyn})}=\arg\bra{\psi_{0}}\mathcal{M}_{N}^{(r_{N})}...\mathcal{M}_{1}^{(r_{1})}\ket{\psi_{0}}-\arg\bra{\psi_{0}}\mathcal{P}_{N}...\mathcal{P}_{1}\ket{\psi_{0}}
\end{equation}
the dynamical component of the measurement-induced phase.

This consideration implies that the averaged phase, $\bar{\chi}$,
too may not be assigned the meaning of a purely geometrical phase.
However, in the case of averaging, our discussion below does not provide
an algorithm for separating the phase into a geometrical and a dynamical
components.

For adiabatic Hamiltonian evolution driven by a Hamiltonian $H(t\in[0,T])$,
the phase factor accumulated in the course of the evolution can be
split into dynamical and geometrical (Berry) components \citep{Berry1984}.
These components present the following property: evolving the system
in the opposite direction, $H(t)\rightarrow H(T-t)$, keeps the dynamical
component unchanged and reverses the sign of the Berry component.
This has been the basis for separating dephasing in open systems undergoing
adiabatic evolution into dynamical and geometrical components \citep{Whitney2003,Whitney2005,Whitney2006,Berger2015,Snizhko2019b,Snizhko2019e}.
Evidently, it is of interest to investigate the behavior of measurement-induced
phases under reversing the measurement sequence.

Consider the same protocol as in Sec.~\ref{sec:IIB_phases_definitions}
but with the intermediate measurements executed in the opposite order.
The postselected phase, $\chi_{\{r_{k}\}}$, cf.~Eq.~(\ref{eq:chi_definition}),
is then defined through $\bra{\psi_{0}}\mathcal{M}_{1}^{(r_{1})}...\mathcal{M}_{N}^{(r_{N})}\ket{\psi_{0}}=\bra{\psi_{0}}\mathcal{M}_{N}^{(r_{N})\dagger}...\mathcal{M}_{1}^{(r_{1})\dagger}\ket{\psi_{0}}^{*}$.
For Hermitian Kraus operators, the last expression is equal to $\bra{\psi_{0}}\mathcal{M}_{N}^{(r_{N})}...\mathcal{M}_{1}^{(r_{1})}\ket{\psi_{0}}^{*}$,
meaning that the phase $\chi_{\{r_{k}\}}$ reverses its sign, while
the probability of the readout sequence $P_{\{r_{k}\}}$ is unchanged,
cf.~Eq.~(\ref{eq:chi_definition}). For general (non-Hermitian)
Kraus operators, however, no simple relation exists between the direct
and the reversed protocols. Moreover, as we show below, even the geometrical
component of the phase $\chi_{\{r_{k}\}}$ does not possess a simple
symmetry with respect to the reversal of the protocol's direction.
One may then define the symmetric and the antisymmetric components
of the measurement-induced phases:

\begin{align}
\chi_{\{r_{k}\}}^{s/a} & =\chi_{\{r_{k}\}}^{(d=+1)}\pm\chi_{\{r_{k}\}}^{(d=-1)},\label{eq:def_phases_sym/antisym_postselected}\\
\bar{\chi}^{s/a} & =\bar{\chi}^{(d=+1)}\pm\bar{\chi}^{(d=-1)},\label{eq:def_phases_sym/antisym_averaged}
\end{align}
where $\chi_{\{r_{k}\}}^{(d)}$ for the direct ($d=+1$) and reversed
($d=-1$) protocols are defined via
\begin{align}
\bra{\psi_{0}}\mathcal{M}_{N}^{(r_{N})}...\mathcal{M}_{1}^{(r_{1})}\ket{\psi_{0}} & =\sqrt{P_{\{r_{k}\}}^{(d=+1)}}e^{i\chi_{\{r_{k}\}}^{(d=+1)}},\label{eq:def_postselected_d+}\\
\bra{\psi_{0}}\mathcal{M}_{1}^{(r_{1})}...\mathcal{M}_{N}^{(r_{N})}\ket{\psi_{0}} & =\sqrt{P_{\{r_{k}\}}^{(d=-1)}}e^{i\chi_{\{r_{k}\}}^{(d=-1)}},\label{eq:def_postselected_d-}
\end{align}
and the averaged phases $\bar{\chi}^{(d)}$ are defined via
\begin{equation}
e^{2i\bar{\chi}^{(d)}-\alpha^{(d)}}=\sum_{\{r_{k}\}}P_{\{r_{k}\}}^{(d)}e^{2i\chi_{\{r_{k}\}}^{(d)}}.\label{eq:def_averaged_d}
\end{equation}
Similarly, we introduce the symmetric and antisymmetric components
of the probabilities and the dephasing parameter:
\begin{align}
P_{\{r_{k}\}}^{s} & =\sqrt{P_{\{r_{k}\}}^{(d=+1)}P_{\{r_{k}\}}^{(d=-1)}},\label{eq:def_prob_sym}\\
P_{\{r_{k}\}}^{a} & =\sqrt{P_{\{r_{k}\}}^{(d=+1)}/P_{\{r_{k}\}}^{(d=-1)}},\label{eq:def_prob_antisym}\\
\alpha^{s/a} & =\alpha^{(d=+1)}\pm\alpha^{(d=-1)}.\label{eq:def_dephasing_sym/antisym}
\end{align}

\begin{figure}
\begin{centering}
\includegraphics[width=0.6\columnwidth]{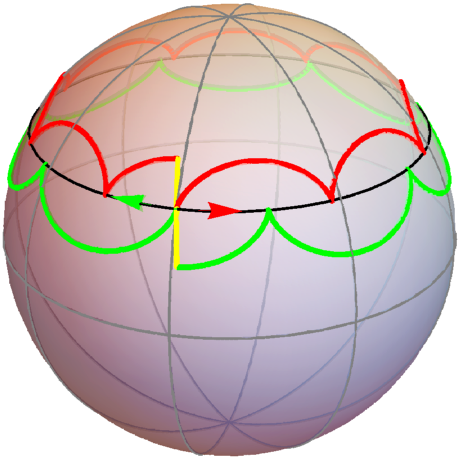}
\par\end{centering}
\caption{\label{fig:trajectory_asymmetry}The state trajectories of the system
for the direct and reversed measurement sequences may be drastically
different. A spin 1/2 system, whose states are represented on the
Bloch sphere, is subject to a sequence of generalized measurements,
pertaining to the spin projections around a given parallel (black
line), in accordance with the protocol described in Sec.~\ref{sec:III_particular_protocol}.
All measurements of either the clockwise (red) or the anticlockwise
(green) sequences are read out and postseleted to have $r_{k}=0$.
The yellow lines connect $\protect\ket{\psi_{N}}$ to $\protect\ket{\psi_{0}}$
by the shortest geodesic on the Bloch sphere. This line represents
closing the trajectories by a postselected projective measurement
at the end of the measurement sequence, cf.~the discussion between
Eqs.~(\ref{eq:def_intermediate_states}--\ref{eq:chi_definition}).
It is known that, similar to the Berry phase, the Pancharatnam phase
(\ref{eq:Pancharatnam_phase}) for a spin 1/2 system may be expressed
in terms of the solid angle enclosed by the closed trajectory. The
trajectories under the direct and reversed measurement sequences do
not subtend the same solid angle on the Bloch sphere. Their solid
angles differ by sign and \emph{in absolute value}. Clearly the Pancharatnam
phases, cf.~Eq.~(\ref{eq:Pancharatnam_phase}), under the two time-reversed
sequences are different.}
\end{figure}

The above considerations lead to the following major conclusion: unlike
in adiabatic Hamiltonian evolution, the classification of contributions
to the measurement-induced phase into symmetric vs. antisymmetric,
does not coincide with the classification into dynamical vs. geometrical
contributions. An intuitive understanding of this result relies on
the following observation: the intermediate states, $\ket{\psi_{k}}$,
for the direct and the reversed measurement sequences form different
trajectories (cf.~Fig.~\ref{fig:trajectory_asymmetry}), implying
that the geometrical phase components (\ref{eq:Pancharatnam_phase})
are different in magnitude for the direct and the reversed protocols.
We present an explicit illustration of this in Sec.~\ref{subsec:IIIE3_Hamiltonian_limit}.

\section{\label{sec:III_particular_protocol}Weak-measurement-induced phases:
an explicit example}

In the rest of this paper we focus on a specific measurement class
(Sec.~\ref{sec:IIIA_detector_model}), and a specific set of measurement
sequences that make the system's spin 1/2 follow closed trajectories
on the Bloch sphere (Sec.~\ref{sec:IIIB_measurement_sequences+scaling}).
In Sec.~\ref{sec:IIIC_postselected_analytics} we present analytic
expressions for the postselected measurement-induced phase under this
protocol. In Sec.~\ref{sec:IIID_averaged_semianalytics}, we outline
a procedure that allows one to calculate the averaged phase in an
efficient manner. Explicit results, pertaining to certain limiting
cases (e.g., nearly projective measurements) are presented in Sec.~\ref{sec:IIIE_limiting cases}.

\subsection{\label{sec:IIIA_detector_model}The measurement model}

Here we describe the measurement model that gives rise to the specific
back-action matrices used throughout the rest of the paper. We consider
a two-state system (with basis states $\ket 0_{s}$ and $\ket 1_{s}$)
and a two-state detector (with basis states $\ket 0_{D}$ and $\ket 1_{D}$)
with the measurement procedure being as follows. The detector is prepared
in state $\ket{D_{i}}=\ket 0_{D}$. We choose the system-detector
interaction Hamiltonian, cf.~Sec.~\ref{sec:IIA_measurement_theory},
to be
\begin{equation}
H_{\mathrm{s-d}}=-\frac{\lambda(t)}{2}\left(1-(\mathbf{n}^{(s)}\cdot\bm{\sigma}^{(s)})\right)(\mathbf{n}^{(D)}\cdot\bm{\sigma}^{(D)}).\label{eq:Hsd_particular}
\end{equation}
It is switched on during a time interval of duration $T$, i.e., $\lambda(t<0)=\lambda(t>T)=0$;
$\bm{\sigma}^{(s/D)}$ are the vectors of Pauli matrices $(\sigma_{x},\sigma_{y},\sigma_{z})$
acting on the system/detector. The vectors
\begin{multline}
\mathrm{\mathbf{n}}^{(s/D)}=(\sin\theta^{(s/D)}\cos\varphi^{(s/D)},\\
\sin\theta^{(s/D)}\sin\varphi^{(s/D)},\cos\theta^{(s/D)})
\end{multline}
determine the system observable measured, ($\mathbf{n}^{(s)}\cdot\bm{\sigma}^{(s)}$),
and the effect of the system-detector interaction on the detector
state. Note that the vectors $\mathbf{n}^{(s)}$ and $\mathbf{n}^{(D)}$
are normalized, $\mathbf{n}^{(s)}\cdot\mathbf{n}^{(s)}=\mathbf{n}^{(D)}\cdot\mathbf{n}^{(D)}=1$.
The arbitrary initial state of the measured system, $\ket{\psi}=a_{s}\ket 0_{s}+b_{s}\ket 1_{s}$,
evolves under the system-detector coupling according to
\begin{multline}
\ket{\psi}\ket{D_{i}}\\
\rightarrow\exp\left[i\frac{g}{2}\left(1-(\mathbf{n}^{(s)}\cdot\bm{\sigma}^{(s)})\right)(\mathbf{n}^{(D)}\cdot\bm{\sigma}^{(D)})\right]\ket{\psi}\ket{D_{i}}\\
=\ket{\psi^{(0)}}\ket 0_{D}+\ket{\psi^{(1)}}\ket 1_{D},
\end{multline}
where $g=\int_{0}^{T}dt\lambda(t)$. After the interaction has been
switched off, $\sigma_{z}^{(D)}$ is measured projectively, yielding
a readout $r\in\{0;1\}$ corresponding to the post-measurement detector
states $\ket r_{D}$. The back-action matrices (representing the Kraus
operators) are thus 
\begin{equation}
\mathcal{M}^{(r)}=R^{-1}(\mathbf{n}^{(s)})M^{(r)}R(\mathrm{\mathbf{n}}^{(s)})\label{eq:our_Kraus_definition}
\end{equation}
 with
\begin{eqnarray}
M^{(0)} & = & \begin{pmatrix}1 & 0\\
0 & \cos g+i\sin g\cos\theta^{(D)}
\end{pmatrix},\label{eq:back_action_0}\\
M^{(1)} & = & \begin{pmatrix}0 & 0\\
0 & i\sin g\sin\theta^{(D)}e^{i\varphi^{(D)}}
\end{pmatrix},\label{eq:back_action_1}\\
R(\mathbf{n}^{(s)}) & = & \begin{pmatrix}\cos\frac{\theta^{(s)}}{2} & \sin\frac{\theta^{(s)}}{2}e^{-i\varphi^{(s)}}\\
\sin\frac{\theta^{(s)}}{2} & -\cos\frac{\theta^{(s)}}{2}e^{-i\varphi^{(s)}}
\end{pmatrix}.\label{eq:rotation}
\end{eqnarray}
When $\mathbf{n}^{(s)}=(0,0,1)$, the matrices $M^{(r)}$ alone determine
the back-action. For a general $\mathbf{n}^{(s)}$, the matrix $R(\mathbf{n}^{(s)})$
induces a unitary rotation: the eigenbasis of $(\mathbf{n}^{(s)}\cdot\bm{\sigma}^{(s)})=R^{-1}(\mathbf{n}^{(s)})\sigma_{z}^{(s)}R(\mathbf{n}^{(s)})$
is given by $R^{-1}(\mathbf{n}^{(s)})\ket{0/1}_{s}$. One thus sees
that the role of $M^{(r)}$ is to determine the back-action in the
eigenbasis of the measured observable $(\mathbf{n}^{(s)}\cdot\bm{\sigma}^{(s)})$.

It is important to understand in detail the evolution of the system
state during the measurement process. Consider the case of $\mathbf{n}^{(s)}=(0,0,1)$.\footnote{For arbitrary $\mathbf{n}^{(s)}$, the effect is the same if considered
in the eigenbasis of $(\mathbf{n}^{(s)}\cdot\bm{\sigma}^{(s)})$.} If the initial state $\ket{\psi}=\ket 0_{s}$, the measurement yields
$r=0$ with probability 1 and the state remains unchanged. For the
initial state $\ket{\psi}=\ket 1_{s}$, the probabilities of the readouts
are $p_{r=0}=1-\sin^{2}g\sin^{2}\theta^{(D)}$ and $p_{r=1}=\sin^{2}g\sin^{2}\theta^{(D)}$;
the state becomes $\ket{\psi^{(r)}}=e^{i\phi_{r}}\ket 1_{s}$ with
a readout-dependent phase $\phi_{r}$. For a generic initial state,
both readouts are possible with some probabilities $p_{r}$, cf.~Eq.~(\ref{eq:readout_probability_general}),
yet the back-action on the state does not reduce to a phase multiplication.
The $r=1$ readout, whose back-action is described by $M^{(1)}$,
projects the state onto $\ket 1_{s}$. For $r=0$ readout, $M^{(0)}$
describes pulling the state towards the north pole on the Bloch sphere
(i.e., closer to $\ket 0_{s}$) and rotating it around the $z$ axis,
cf.~Fig.~\ref{fig:measurement_back_action}.

\begin{figure}
\begin{centering}
\includegraphics[width=0.8\columnwidth]{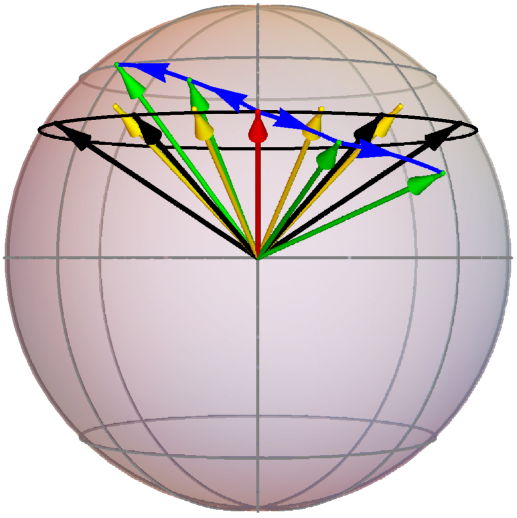}
\par\end{centering}
\caption{\label{fig:trajectory_asymmetry_explanation}Emergence of asymmetry
of time-reversed measurement-induced trajectories. Two measurement
sequences of spin pojections $(\mathbf{n}^{(s)}\cdot\bm{\sigma}^{(s)})$
onto subsequent directions $\mathbf{n}^{(s)}$ located on the same
parallel (black line) are performed, a clockwise and an anticlockwise
one respectively. We illustrate the effect of the first two measurements
of each sequence (clockwise and anti-clockwise). All measurements
are assumed to yield readouts $r=0$ and are characterized by Kraus
operators $\mathcal{M}^{(0)}$ (\ref{eq:our_Kraus_definition}). The
initial state is represented by the red arrow, and the measurement
axes are shown by black arrows. If the back-action operator $\mathcal{M}^{(0)}$
is Hermitian, the clockwise and the counterclockwise trajectories
are mirror reflections of each other (yellow arrows). However, for
non-Hermitian back-action, there is an extra rotation around the measurement
axes, cf.~Fig.~\ref{fig:measurement_back_action}, leading to the
two trajectories not being simply related to each other (green arrows
show the state locations after the respective measurements; these
locations are connected by the blue arrows to show what is the clockwise
trajectory and what is the anticlockwise one).}
\end{figure}

This rotation is a key feature of weak measurement and is absent in
the case of projective measurements. Indeed, in a projective measurement,
the $r=0$ readout would imply the final state $\ket 0_{s}$, and
any rotation around the $z$ axis would become insignificant. Note
that this rotation only happens when $M^{(0)}$ has an imaginary component,
i.e., when $\mathcal{M}^{(0)}$ is non-Hermitian. The idiosyncrasy
of Hermitian back-action matrices has been discussed in Sec.~\ref{sec:IIC_dyn/geom+sym/antisym_classifications}.
It is this non-Hermitian back-action that gives rise to asymmetric
state trajectories as shown in Fig.~\ref{fig:trajectory_asymmetry},
cf.~Fig.~\ref{fig:trajectory_asymmetry_explanation} for a detailed
explanation.

We note in passing that the back-action matrices in Eqs.~(\ref{eq:back_action_0}--\ref{eq:back_action_1})
can appear in a wider context than the toy detector model introduced
here\textcolor{blue}{.} In some contexts, measurements with such back-action
are known as partial or interaction-free measurements \citep{Elitzur2001,Paraoanu2006,Xu2011}
and can be implemented by optical means \citep{Elitzur2001,Xu2011}
or in superconducting qubits \citep{Paraoanu2006}. A particularly
natural setting for such back-action is presented by imperfect optical
polarizers. A polarizer is a detector in the sense that the light
can pass through the polarizer (identified with $r=0$) or not pass
(be absorbed, reflected etc., $r=1$). An ideal polarizer, letting
one polarization through and completely reflecting the other, is equivalent
to a strong measurement being applied to a photon. A non-ideal polarizer,
letting one polarization through completely, while the other polarization
is partially reflected and partially transmitted, can be described
as a measurement with back-action matrices of the form (\ref{eq:back_action_0}--\ref{eq:back_action_1}).
Therefore, our proposed protocols, detailed below, may, in principle,
be implemented in a variety of experimental settings.

\subsection{\label{sec:IIIB_measurement_sequences+scaling}The measurement sequences
and the scaling limit}

Hereafter we focus on studying a specific family of measurement sequences.
We consider the system initial state 
\begin{equation}
\ket{\psi_{0}}=\cos\frac{\theta}{2}\ket 0_{s}+\sin\frac{\theta}{2}\ket 1_{s}.\label{eq:psi_0}
\end{equation}
We choose our measurements to be associated with the measurement axes
\begin{equation}
\mathbf{n}_{k}^{(s)}=(\sin\theta_{k}^{(s)}\cos\varphi_{k}^{(s)},\sin\theta_{k}^{(s)}\sin\varphi_{k}^{(s)},\cos\theta_{k}^{(s)}),\label{eq:n_axes_sequence}
\end{equation}
 with 
\begin{equation}
(\theta_{k},\varphi_{k})=(\theta,2\pi kd/(N+1)),\label{eq:(theta,phi)_axes_sequence}
\end{equation}
i.e., all the measurement axes belong to a particular parallel corresponding
to the polar angle $\theta$, and $d=\pm1$ denotes whether the sequence
is performed clockwise/counterclockwise. We fix the measurement parameters
$g_{k}$ and $\mathbf{n}_{k}^{(D)}$ to be $g$ and $\mathbf{n}^{(D)}=(\sin\theta^{(D)}\cos\varphi^{(D)},\sin\theta^{(D)}\sin\varphi^{(D)},\cos\theta^{(D)})$,
independently of the measurement number $k$. For simplicity, we put
$\varphi^{(D)}=-\pi/2$.

We will be interested in the limit $N\rightarrow\infty$, where the
measurement sequence becomes quasicontinuous. If one keeps $g$ and
$\theta^{(D)}$ constant when taking the $N\rightarrow\infty$ limit,
a sequence of infinite number of finite strength measurements becomes
equivalent to a sequence of projective measurements and yields the
Pancharatnam phase, cf.~Appendix~\ref{sec:appendix_scaling_regimes}.
In order to avoid this trivial limiting case, one needs to scale $g$
and $\theta^{(D)}$ with $N$. In Appendix~\ref{sec:appendix_scaling_regimes},
we show that among the large number of possible approaches to the
continuum limit, there is a unique scaling procedure that avoids a
trivial limit.

This non-trivial scaling procedure corresponds to $g=\sqrt{4C/N}$
and $\theta^{(D)}=\pi/2+A/\sqrt{CN}$ with the parameters $C\geq0$
and $A\in\mathbb{R}$. With such scaling, the back-action matrices
in Eqs.~(\ref{eq:back_action_0}--\ref{eq:back_action_1}) become
\begin{align}
M^{(0)} & =\begin{pmatrix}1 & 0\\
0 & \exp\left(-2\frac{C+iA}{N}\right)+O\left(\frac{1}{N^{2}}\right)
\end{pmatrix},\label{eq:back_action_0_scaling}\\
M^{(1)} & =\begin{pmatrix}0 & 0\\
0 & \sqrt{\frac{4C}{N}}+O\left(\frac{1}{N^{3/2}}\right)
\end{pmatrix}.\label{eq:back_action_1_scaling}
\end{align}
The parameter $C$ controls the measurement strength (how much the
state is pulled towards the measurement axis for the $r=0$ readout),
while $A\in\mathbb{R}$ controls the non-Hermiticity of $M^{(0)}$
(and $\mathcal{M}^{(0)}$ in Eq.~(\ref{eq:our_Kraus_definition})).
Since non-Hermiticity is the cause of asymmetric behavior (as was
shown in Secs.~\ref{sec:IIC_dyn/geom+sym/antisym_classifications},
\ref{sec:IIIA_detector_model}), we call $A$ the asymmetry parameter.

The non-Hermitian contribution to the measurement back-action can
be interpreted as Hamiltonian evolution:
\begin{multline}
M^{(0)}=\begin{pmatrix}1 & 0\\
0 & \exp\left(-2\frac{C+iA}{N}\right)
\end{pmatrix}\\
=\begin{pmatrix}1 & 0\\
0 & \exp\left(-2\frac{C}{N}\right)
\end{pmatrix}\exp\left(-iH\Delta t\right),\label{eq:asymmetry_as_Hamiltonian_evolution}
\end{multline}
where $\Delta t=N^{-1}$ and $H=A\left(\mathbb{I}-\sigma_{z}^{(s)}\right)$.
Therefore, this back-action could, in principle, arise as a result
of a measurement with Hermitian back-action applied to a system evolving
under the Hamiltonian $H$. This, however, is not how the back-action
emerges here: the system does not have its own Hamiltonian, nor does
the detector model have any term in the Hamiltonian (\ref{eq:Hsd_particular})
acting solely on the system. Nevertheless, Eq.~(\ref{eq:asymmetry_as_Hamiltonian_evolution})
shows that \emph{for the purposes of investigating the effect on the
system state}, the measurements we consider are equivalent to measurements
with a Hermitian back-action (determined by $C$) supplemented with
Hamiltonian evolution of the system (determined by $A$). We find
this equivalence useful for connecting our results to the known results
for Hamiltonian-evolution-induced phase factors in Sec.~\ref{subsec:IIIE3_Hamiltonian_limit}.\footnote{Note also that the scaling of the back-action matrix in Eq.~(\ref{eq:back_action_0_scaling})
is the ``natural'' one in the following case: the measurements are
implemented with polarizers, where the degree of polarization is determined
by the polarizer thickness. Indeed, for such a polarizer, the degree
of letting the ``wrong'' polarization through would drop exponentially
with the thickness $L$ of the polarizer. At the same time, different
refraction indices for the two polarizations would also result in
a phase difference proportional to $L$. Adjusting the polarizer thickness
$L\sim N^{-1}$ according to the number N of measurements employed
would result in the back-action given in Eq.~(\ref{eq:back_action_0_scaling}),
applied to the polarization of the transmitted light. This should
enable a relatively easy check of our predictions concerning the case
when all the measurements are postselected to yield $r_{k}=0$, cf.
Secs.~\ref{sec:IIIC_postselected_analytics}, \ref{sec:IV_top_trans_postselected},
\ref{subsec:VI.A_interferometric_detection_schemes}.}

\subsection{\label{sec:IIIC_postselected_analytics} Measurement-induced phase
in postselected measurement sequences}

Here we investigate the behavior of the postselected phase, $\chi_{\left\{ r_{k}\right\} }^{(d=\pm1)}$
defined in Sec.II, Eqs.~(\ref{eq:def_postselected_d+}, \ref{eq:def_postselected_d-}).
We focus on a specific readout sequence in which all detector readouts
are $r_{k}=0$. Such a choice is based on the following observation.
Within the measurement model described in Sec.~\ref{sec:IIIA_detector_model},
$r=0$ readout implies that the detector state before a measurement
coincides with the detector state after the measurement. This allows
for designing a simple observation scheme for $\chi_{\left\{ r_{k}=0\right\} }^{(d)}$,
as described in Sec.~\ref{sec:VI_experimental_implementation}.

The parameter $d=\pm1$ denotes the direction of the measurement sequence,
cf.~Eq.~(\ref{eq:(theta,phi)_axes_sequence}). We next calculate
$\chi_{\left\{ r_{k}=0\right\} }^{(d)}$ for both directions, keeping
$d$ unspecified. Using Eqs.~(\ref{eq:our_Kraus_definition}--\ref{eq:rotation})
and the explicit definitions for the initial state $\ket{\psi_{0}}$
(\ref{eq:psi_0}), the measurement axes $\mathrm{\mathbf{n}}_{k}^{(s)}$
(\ref{eq:n_axes_sequence}), and the protocol direction $d$ (\ref{eq:(theta,phi)_axes_sequence}),
one shows that
\begin{multline}
\sqrt{P_{\{r_{k}=0\}}^{(d)}}e^{i\chi_{\{r_{k}=0\}}^{(d)}}=\bra{\psi_{0}}\mathcal{M}_{N}^{(0)}...\mathcal{M}_{1}^{(0)}\ket{\psi_{0}}\\
=\begin{pmatrix}1 & 0\end{pmatrix}\delta R^{(d)}(M^{(0)}\delta R^{(d)})^{N}\begin{pmatrix}1\\
0
\end{pmatrix},\label{eq:simplification_postselected}
\end{multline}
where \begin{widetext}
\begin{equation}
\delta R^{(d)}=R(\mathrm{\mathbf{n}}_{k}^{(s)})R^{-1}(\mathbf{n}_{k-1}^{(s)})=\begin{pmatrix}\cos^{2}\frac{\theta}{2}+\sin^{2}\frac{\theta}{2}\exp\left(-\frac{2\pi id}{N+1}\right) & \frac{1}{2}\left[1-\exp\left(-\frac{2\pi id}{N+1}\right)\right]\sin\theta\\
\frac{1}{2}\left[1-\exp\left(-\frac{2\pi id}{N+1}\right)\right]\sin\theta & \sin^{2}\frac{\theta}{2}+\cos^{2}\frac{\theta}{2}\exp\left(-\frac{2\pi id}{N+1}\right)
\end{pmatrix}.\label{eq:deltaR}
\end{equation}
Using Eq.~(\ref{eq:back_action_0_scaling}), diagonalizing $M^{(0)}\delta R^{(d)}$,
and taking the limit of $N\rightarrow\infty$, one finds that
\begin{equation}
\sqrt{P_{\{r_{k}=0\}}^{(d)}}e^{i\chi_{\{r_{k}=0\}}^{(d)}}=e^{i\pi d(\cos\theta-1)-Z}\left(\cosh\tau+Z\frac{\sinh\tau}{\tau}\right),\label{eq:postselected_result}
\end{equation}
\end{widetext}where $Z=C+iA+i\pi d\cos\theta$ and $\tau=\sqrt{Z^{2}-\pi^{2}\sin^{2}\theta}$.
Note that the definition of $\tau$ through the square root allows
for a sign ambiguity. Since Eq.~(\ref{eq:postselected_result}) is
symmetric under $\tau\rightarrow-\tau$, the actual sign does not
matter and one can choose any convention for calculating the square
root. Note also that the prefactor $e^{i\pi d(\cos\theta-1)}$ is
exactly the Pancharatnam phase of the system subjected to a quasicontinuous
sequence of projective measurements along the parallel corresponding
to $\theta$.

The r.h.s. of Eq.~(\ref{eq:postselected_result}) obeys a number
of symmetries. First, the expression is invariant under simultaneous
replacement of $d\rightarrow-d$ and $\theta\rightarrow\pi-\theta$.
Second, the expression remains unaffected under $d\rightarrow-d$
and $A\rightarrow-A$ accompanied by the complex conjugation. From
the latter, it follows that for $A=0$, $P_{\{r_{k}=0\}}^{(d=+1)}=P_{\{r_{k}=0\}}^{(d=-1)}$
and $\chi_{\{r_{k}=0\}}^{(d=+1)}=-\chi_{\{r_{k}=0\}}^{(d=-1)}(\mathrm{mod}\,2\pi)$.
That is, at $A=0$ the probability only has a non-trivial symmetric
component, and the phase only has the antisymmetric component. Away
from $A=0$, the phase and the postselection probability, each has
both the symmetric and the antisymmetric components (\ref{eq:def_phases_sym/antisym_postselected},
\ref{eq:def_prob_sym}, \ref{eq:def_prob_antisym}).

\subsection{\label{sec:IIID_averaged_semianalytics}How to calculate the averaged
phase}

Here we derive a relatively simple expression for the averaged measurement-induced
phase, $\bar{\chi}^{(d)}$ in Eq.~(\ref{eq:def_averaged_d}). While
our result does not constitute a fully analytical expression for $\bar{\chi}^{(d)}$,
it facilitates general analysis and efficient numerical study of the
averaged phase behavior.

Note that similarly to Eq.~(\ref{eq:simplification_postselected}),
for an arbitrary readout sequence $\left\{ r_{k}\right\} $,
\begin{multline}
\bra{\psi_{0}}\mathcal{M}_{N}^{(r_{N})}...\mathcal{M}_{1}^{(r_{1})}\ket{\psi_{0}}\\
=\begin{pmatrix}1 & 0\end{pmatrix}\delta R^{(d)}M^{(r_{N})}\delta R^{(d)}...\delta R^{(d)}M^{(r_{1})}\delta R^{(d)}\begin{pmatrix}1\\
0
\end{pmatrix}
\end{multline}
with $M^{(r_{k})}$ defined in Eqs.~(\ref{eq:back_action_0_scaling}--\ref{eq:back_action_1_scaling})
and $\delta R^{(d)}$ defined in Eq.~(\ref{eq:deltaR}). Then
\begin{multline}
e^{2i\bar{\chi}^{(d)}-\alpha^{(d)}}=\sum_{\{r_{k}\}}\left(\bra{\psi_{0}}\mathcal{M}_{N}^{(r_{N})}...\mathcal{M}_{1}^{(r_{1})}\ket{\psi_{0}}\right)^{2}\\
=\sum_{\{r_{k}\}}\begin{pmatrix}1\\
0
\end{pmatrix}^{T}\otimes\begin{pmatrix}1\\
0
\end{pmatrix}^{T}\delta R_{4}^{(d)}M_{4}^{(r_{N})}...M_{4}^{(r_{1})}\delta R_{4}^{(d)}\begin{pmatrix}1\\
0
\end{pmatrix}\otimes\begin{pmatrix}1\\
0
\end{pmatrix}\\
=\begin{pmatrix}1\\
0
\end{pmatrix}^{T}\otimes\begin{pmatrix}1\\
0
\end{pmatrix}^{T}\delta R_{4}\left(\mathfrak{M}^{(d)}\right)^{N}\begin{pmatrix}1\\
0
\end{pmatrix}\otimes\begin{pmatrix}1\\
0
\end{pmatrix},
\end{multline}
where $\delta R_{4}^{(d)}=\delta R^{(d)}\otimes\delta R^{(d)}$, $M_{4}^{(r_{k})}=M^{(r_{k})}\otimes M^{(r_{k})}$,
$\otimes$ denotes the tensor product, and $\mathfrak{M}^{(d)}=\sum_{r}M_{4}^{(r)}\delta R_{4}^{(d)}$.
Therefore,
\begin{multline}
e^{2i\bar{\chi}^{(d)}-\alpha^{(d)}}=\begin{pmatrix}1 & 0 & 0 & 0\end{pmatrix}\delta R_{4}^{(d)}\left(\mathfrak{M}^{(d)}\right)^{N}\begin{pmatrix}1 & 0 & 0 & 0\end{pmatrix}^{T}\\
=\left[\delta R_{4}^{(d)}\left(\mathfrak{M}^{(d)}\right)^{N}\right]_{11},\label{eq:simplification_averaged}
\end{multline}
where\begin{widetext}
\begin{equation}
\mathfrak{M}^{(d)}=\begin{pmatrix}1+\frac{2i\pi d\cos\theta}{N} & -\frac{i\pi d\sin\theta}{N} & -\frac{i\pi d\sin\theta}{N} & 0\\
-\frac{i\pi d\sin\theta}{N} & 1-2\frac{C+iA}{N} & 0 & -\frac{i\pi d\sin\theta}{N}\\
-\frac{i\pi d\sin\theta}{N} & 0 & 1-2\frac{C+iA}{N} & -\frac{i\pi d\sin\theta}{N}\\
0 & -\frac{i\pi d\sin\theta}{N} & -\frac{i\pi d\sin\theta}{N} & 1-\frac{2i\pi d\cos\theta}{N}-\frac{4iA}{N}
\end{pmatrix}+O\left(\frac{1}{N^{2}}\right).\label{eq:MdeltaR_4x4}
\end{equation}
\end{widetext}

What enabled a fully analytical calculation in Sec.~\ref{sec:IIIC_postselected_analytics}
is the possibility to diagonalize $M^{(0)}\delta R^{(d)}$ analytically.
Here, diagonalizing $\mathfrak{M}^{(d)}$ analytically is a formidable
task. However, it can be diagonalised numerically. Suppose one diagonalised
$\mathfrak{M}^{(d)}$,
\begin{equation}
\mathfrak{M}^{(d)}=VDV^{-1}
\end{equation}
with $D=\mathrm{diag}(\lambda_{1},\lambda_{2},\lambda_{3},\lambda_{4})$
and $\lambda_{j}=1+x_{j}/N+O(N^{-2})$. Then in the limit $N\rightarrow\infty$,
\begin{equation}
e^{2i\bar{\chi}^{(d)}-\alpha^{(d)}}=\left[\left(\mathfrak{M}^{(d)}\right)^{N}\right]_{11}=\left[V\mathcal{D}V^{-1}\right]_{11},\label{eq:simplification_averaged_largeN}
\end{equation}
with $\mathcal{D}=\mathrm{diag}(e^{x_{1}},e^{x_{2}},e^{x_{3}},e^{x_{4}})$.

Expressions (\ref{eq:MdeltaR_4x4}, \ref{eq:simplification_averaged_largeN})
not only provide means for efficient numeric calculation of the averaged
phase, they also allow one to make some analytic conclusions. Namely,
one can show that the averaged phase obeys the same symmetries as
the postselected phase (cf.~Sec.~\ref{sec:IIIC_postselected_analytics}).
Observe that $\left.\mathfrak{M}^{(d)}\right|_{\theta\rightarrow\pi-\theta}=U^{-1}\mathfrak{M}^{(-d)}U$,
where $U=\mathrm{diag}(1,-1,-1,1)$. Therefore, $e^{2i\bar{\chi}^{(d)}-\alpha^{(d)}}$
remains invariant under simultaneous replacement $d\rightarrow-d$,
$\theta\rightarrow\pi-\theta$:
\begin{multline}
\left.e^{2i\bar{\chi}^{(-d)}-\alpha^{(-d)}}\right|_{\theta\rightarrow\pi-\theta}=\lim_{N\rightarrow\infty}\left[\left(\left.\mathfrak{M}^{(-d)}\right|_{\theta\rightarrow\pi-\theta}\right)^{N}\right]_{11}\\
=\lim_{N\rightarrow\infty}\left[\left(U^{-1}\mathfrak{M}^{(d)}U\right)^{N}\right]_{11}=\lim_{N\rightarrow\infty}\left[U^{-1}\left(\mathfrak{M}^{(d)}\right)^{N}U\right]_{11}\\
=\lim_{N\rightarrow\infty}\left[\left(\mathfrak{M}^{(d)}\right)^{N}\right]_{11}=e^{2i\bar{\chi}^{(d)}-\alpha^{(d)}}.
\end{multline}
Further, $\left.\mathfrak{M}^{(-d)}\right|_{A\rightarrow-A}=\left(\mathfrak{M}^{(d)}\right)^{*}$,
implying that $e^{2i\bar{\chi}^{(d)}-\alpha^{(d)}}$ is invariant
under applying complex conjugation and simultaneously replacing $d\rightarrow-d$,
$A\rightarrow-A$.

\subsection{\label{sec:IIIE_limiting cases}Limiting cases}

The analytic results of Secs.~\ref{sec:IIIC_postselected_analytics}
and \ref{sec:IIID_averaged_semianalytics} allow one to analyze the
behavior of the postselected, Eq.~(\ref{eq:postselected_result}),
and averaged, Eq.~(\ref{eq:simplification_averaged_largeN}), phases
in a number of limiting cases. In this subsection we discuss three
limiting cases corresponding to $A\rightarrow\infty$, $C\rightarrow\infty$,
and $C=0$.

\subsubsection{\label{subsec:IIIE1_Berry_limit}$A\rightarrow\infty$}

We start with the simplest limiting case of $A\rightarrow\infty$.
This means that the back-action of $r=0$ readouts strongly rotates
the system state around the measurement axis, cf.~Eq.~(\ref{eq:back_action_0_scaling}).
This regime is equivalent to an almost adiabatic Hamiltonian evolution
supplemented by measurements that have a small effect (cf.~Eq.(\ref{eq:asymmetry_as_Hamiltonian_evolution})
in Sec.~\ref{sec:IIIB_measurement_sequences+scaling}). Consequently,
one expects the measurement-induced phase in this limit to coincide
with the Berry phase $\pi d(\cos\theta-1)$ up to small corrections.
This indeed turns out to be the case. For the postselective protocol,
we expand the logarithm of Eq.~(\ref{eq:postselected_result}) at
large $A$. For the averaging protocol, we perform the diagonalization
of Eq.~(\ref{eq:MdeltaR_4x4}) approximately at $A\rightarrow\infty$,
after which we use Eq.~(\ref{eq:simplification_averaged_largeN}).
In both cases, we obtain\begin{widetext}
\begin{equation}
\chi_{\{r_{k}=0\}}^{(d)}=\bar{\chi}^{(d)}=\pi d(\cos\theta-1)+\frac{\pi^{2}\sin^{2}\theta}{2A}-\frac{\pi^{2}\sin^{2}\theta}{4A^{2}}\left[e^{-2C}\sin\left(2A+2\pi d\cos\theta\right)-2\pi d\cos\theta\right]+O(A^{-3}),\label{eq:phases_largeA}
\end{equation}
\begin{equation}
P_{\{r_{k}=0\}}^{(d)}=e^{-\alpha^{(d)}}=\exp\left(-\frac{\pi^{2}\sin^{2}\theta}{2A^{2}}\left[1+2C-e^{-2C}\cos(2A+2\pi d\cos\theta)\right]+O(A^{-3})\right).\label{eq:prob=00003Ddephasing_largeA}
\end{equation}
\end{widetext}It is noteworthy that the results for the postselective
and for the averaging protocols coincide as the $r_{k}\neq0$ readouts
have negligible probability. At higher orders in $A^{-1}$, this is
no longer so.

Note that the phases, $\chi_{\{r_{k}=0\}}^{(d)}$ and $\bar{\chi}^{(d)}$,
do not possess a definite symmetry under $d\rightarrow-d$. In other
words, they feature both symmetric and antisymmetric components, in
agreement with the symmetry-based analysis in Secs.~\ref{sec:IIIC_postselected_analytics},
\ref{sec:IIID_averaged_semianalytics}. The same applies to the postselection
probability, $P_{\{r_{k}=0\}}^{(d)}$, and the dephasing factor, $e^{-\alpha^{(d)}}$.

\subsubsection{\label{subsec:IIIE2_Pancharatnam_limit}$C\rightarrow\infty$}

The limit of $C\rightarrow\infty$ corresponds to almost projective
measurements. Here one expects the induced phase to be the Pancharatnam
phase $\pi d(\cos\theta-1)$ up to small corrections. For the postselective
protocol, expanding Eq.~(\ref{eq:postselected_result}), we find
\begin{equation}
\chi_{\{r_{k}=0\}}^{(d)}=\pi d(\cos\theta-1)+\frac{\pi^{2}\sin^{2}\theta}{2C^{2}}\left[A+\pi d\cos\theta\right]+O(C^{-3}),\label{eq:phase_postselected_largeC}
\end{equation}
\begin{equation}
P_{\{r_{k}=0\}}^{(d)}=\exp\left(-\frac{\pi^{2}\sin^{2}\theta}{C}\left[1-\frac{1}{2C}\right]+O(C^{-3})\right).\label{eq:prob_postselected_largeC}
\end{equation}
For the averaging protocol, we find\begin{widetext}
\begin{equation}
\bar{\chi}^{(d)}=\pi d(\cos\theta-1)+\frac{\pi^{2}\sin^{2}\theta}{2C^{2}}\left[A+\pi d\cos\theta-\pi^{2}\sin^{2}\theta\frac{\sin(4A+4\pi d\cos\theta)-4(A+\pi d\cos\theta)}{16(A+\pi d\cos\theta)^{2}}\right]+O(C^{-3}),\label{eq:phase_averaged_largeC}
\end{equation}
\begin{equation}
e^{-\alpha^{(d)}}=\exp\left(-\frac{\pi^{2}\sin^{2}\theta}{C}\left[1-\frac{1}{2C}\right]+\frac{\pi^{4}\sin^{4}\theta}{2C^{2}}\left[\frac{\sin(2A+2\pi d\cos\theta)}{2(A+\pi d\cos\theta)}\right]^{2}+O(C^{-3})\right).\label{eq:dephasing_averaged_largeC}
\end{equation}
\end{widetext}Note that the asymmetry with respect to $d\rightarrow-d$
is present in the postselected phase but not in the postselection
probability (where it only appears in terms $\propto C^{-3}$). At
the same time, the asymmetry does appear in $\alpha^{(d)}$ at this
order, showing the non-trivial effect of averaging.

When $C\rightarrow\infty$, one recovers the limit of projective measurements,
implying that the resulting phase, $\chi_{\{r_{k}=0\}}^{(d)}$, is
the Pancharatnam phase. Since the postselection probability $P_{\{r_{k}=0\}}^{(d)}=1$,
other readout sequences cannot occur, and the averaged phase is the
same as the postselected one. At large but finite values of $C$,
the two phases are different. One can clearly see the separation of
the $\{r_{k}=0\}$ contribution from that of all the other readout
sequences in both the phases and the postselection probability/dephasing
factor. Remarkably, the other sequences contribute only at $O(C^{-2})$.

\subsubsection{\label{subsec:IIIE3_Hamiltonian_limit}$C=0$}

This limit corresponds to zero strength measurement. The measurements
always yield $r=0$ readouts, and the corresponding back-action (\ref{eq:back_action_0_scaling})
is equivalent to a Hamiltonian evolution, cf.~Eq.~(\ref{eq:asymmetry_as_Hamiltonian_evolution}).
On one hand, this can still be interpreted as the behavior under very
weak measurements. On the other hand, this limit can be understood
as non-adiabatic Hamiltonian evolution and treated within the framework
of Aharonov--Anandan phases \citep{Aharonov1987}. As we show below,
the two treatments yield identical results.

The answer in this limit immediately follows from Eq.~(\ref{eq:postselected_result}),
which yields
\begin{multline}
\sqrt{P_{\{r_{k}=0\}}^{(d)}}e^{i\chi_{\{r_{k}=0\}}^{(d)}}=e^{2i\bar{\chi}^{(d)}-\alpha^{(d)}}\\
=-e^{-iA}\left(\cos\zeta+Z\frac{\sin\zeta}{\zeta}\right),
\end{multline}
where $Z=iA+i\pi d\cos\theta$ and $\zeta=\sqrt{\left(A+\pi d\cos\theta\right)^{2}+\pi^{2}\sin^{2}\theta}$.
Using the same technique as in Sec.~\ref{sec:IIIC_postselected_analytics},
it is possible to obtain the analytic form of the geometrical component
of the phase (\ref{eq:Pancharatnam_phase}):
\begin{multline}
\arg\bra{\psi_{0}}\mathcal{P}_{N}...\mathcal{P}_{1}\ket{\psi_{0}}\\
=\arg\biggl[-e^{-iA}\left(\cos\zeta+Z\frac{\sin\zeta}{\zeta}\right)\\
\times\exp\left(i\frac{A\pi^{2}\sin^{2}\theta}{\zeta^{2}}\left\{ 1-\frac{\sin2\zeta}{2\zeta}\right\} \right)\biggr].
\end{multline}
The dynamical part of the phase is thus 
\begin{multline}
\chi_{\{r_{k}=0\}}^{(d)}-\arg\bra{\psi_{0}}\mathcal{P}_{N}...\mathcal{P}_{1}\ket{\psi_{0}}\\
=-\frac{A\pi^{2}\sin^{2}\theta}{\zeta^{2}}\left\{ 1-\frac{\sin2\zeta}{2\zeta}\right\} .\label{eq:dynamic_phase_no_measurement}
\end{multline}
One sees that neither the dynamical, nor the geometrical part of the
phase possesses a definite symmetry under $d\rightarrow-d$. Each
has both a symmetric and an antisymmestric component.

In the present case, $C=0$, separation into the dynamical and geometrical
components can be obtained following Aharonov and Anandan \citep{Aharonov1987}.
Indeed, the measurement back-action can be interpreted as Hamiltonian
evolution:
\begin{align}
\mathcal{M}_{k}^{(0)}= & \exp\left(-iH_{k}\Delta t\right),\quad\mathcal{M}_{k}^{(1)}=0;\\
H_{k}= & A\left(\mathbb{I}-\mathbf{n}_{k}^{(s)}\cdot\bm{\sigma}^{(s)}\right),\quad\Delta t=N^{-1},
\end{align}
cf.~Eqs.~(\ref{eq:our_Kraus_definition}--\ref{eq:rotation}, \ref{eq:back_action_0_scaling}--\ref{eq:asymmetry_as_Hamiltonian_evolution}).
Then the dynamical phase $-\sum_{k=0}^{N-1}\bra{\psi_{k}}H_{k+1}\ket{\psi_{k}}\Delta t$
in the limit $N\rightarrow\infty$ is given exactly by the r.h.s.
of Eq.~(\ref{eq:dynamic_phase_no_measurement}). This demonstrates
consistency between our definition of the geometrical and dynamical
components of measurement-induced phases and the conventional definition
for the phases induced by Hamiltonian evolution. Further investigation
of the separation of the measurement-induced phases into dynamical
and geometrical components is left for future work.

\section{\label{sec:IV_top_trans_postselected}Topological transitions in
the postselective protocol}

In this section we investigate topological transitions concerning
the postselected phase $\chi_{\{r_{k}=0\}}^{(d)}$. We study these
transitions in Sec.~\ref{sec:IVA_postsel_transition_essence} and
discuss the resulting ``phase diagram'' in the space of measurement
parameters in Sec.~\ref{sec:IVB_postsel_crit_line}.

\subsection{\label{sec:IVA_postsel_transition_essence}The essence of the transitions}

\begin{figure}
\begin{centering}
\includegraphics[width=1\columnwidth]{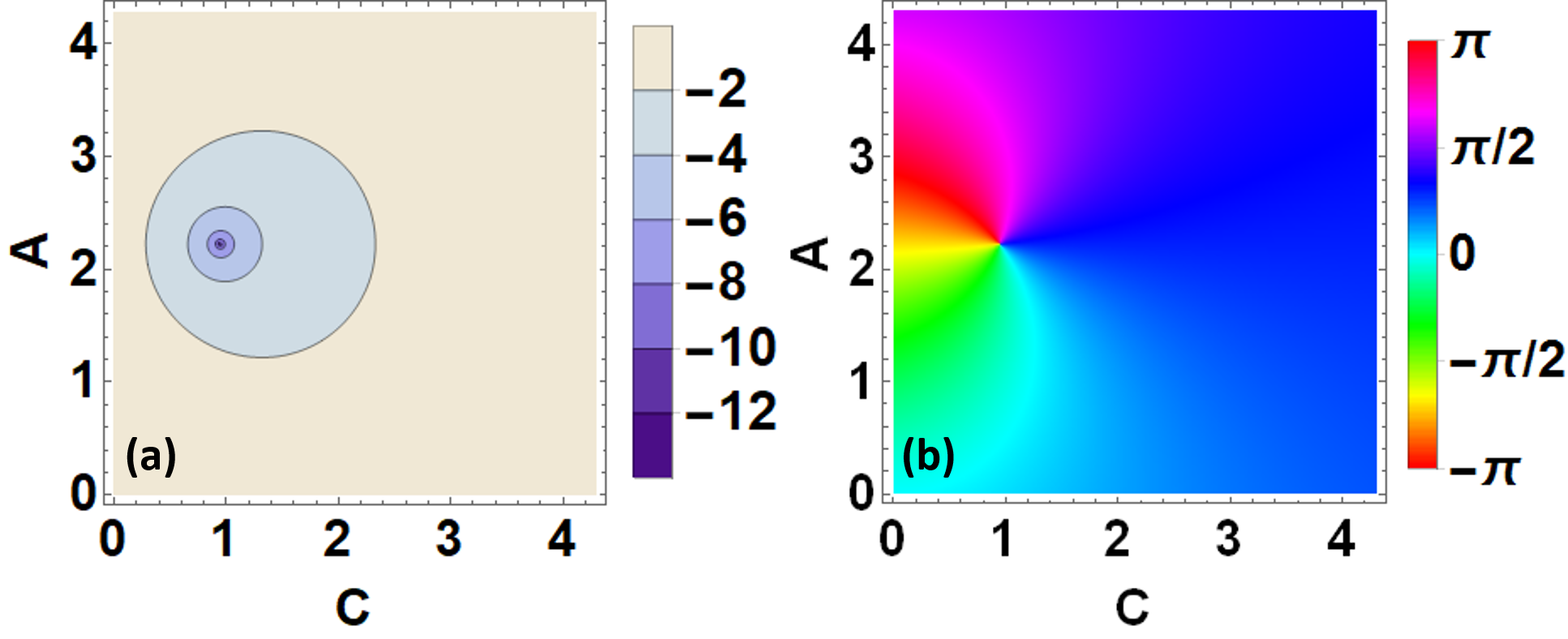}
\par\end{centering}
\caption{\label{fig:logP_divergences_postselected}Vanishing of the probability
and windings of the phase in the postselective protocol, $\left\{ r_{k}=0\right\} $.
(a)---Contour plot of the logarithm of the postselection probability,
$\ln P_{\{r_{k}=0\}}^{(d=+1)}(\theta=3\pi/4)$, (value indicated by
color) as a function of measurement parameters $C$ and $A$, cf.~Eq.~(\ref{eq:postselected_result}).
Note the divergence at $(C,A)\approx(1,2)$. (b)---Dependence of
the phase $\chi_{\{r_{k}=0\}}^{(d=+1)}(\theta=3\pi/4)$ (value indicated
by color) on $C$ and $A$. The phase is ill-defined at the singularity
point $(C,A)\approx(1,2)$. Following the phase value around the singular
point, the phase varies continuously from $-\pi$ to $\pi$, i.e.,
makes a $2\pi$-winding.}
\end{figure}

Consider the postselection probability $P_{\{r_{k}=0\}}^{(d)}$, cf.~Eqs.~(\ref{eq:def_postselected_d+},
\ref{eq:def_postselected_d-}, \ref{eq:postselected_result}), when
the protocol is executed at, for example, $\theta=3\pi/4$, cf.~Fig.~\ref{fig:logP_divergences_postselected}(a).
We note that $\ln P_{\{r_{k}=0\}}^{(d=+1)}(\theta=3\pi/4)$, is bounded
throughout the entire parameter space except for a diveregence near
$C=1$, $A=2$, indicating that $P_{\{r_{k}=0\}}^{(d=+1)}\rightarrow0$
at this special point. This is accompanied by a prominent feature
in the behavior of $\chi_{\{r_{k}=0\}}^{(d=+1)}(\theta=3\pi/4)$,
cf.~Fig.~\ref{fig:logP_divergences_postselected}(b): the phase
$\chi_{\{r_{k}=0\}}^{(d=+1)}(\theta=3\pi/4)$ is ill-defined at the
singularity and makes a $2\pi$-winding around the singular point.
This is a topological feature in the sense that it cannot be eliminated
by a smooth deformation of $\chi_{\{r_{k}=0\}}^{(d=+1)}(\theta=3\pi/4)$
as a function of $(C,A)$. Similar features emerge at other values
of $\theta$, with different locations of the special point in the
$(C,A)$ plane.

For an arbitrary $\theta$, the presence of a phase winding implies
that the phase is ill-defined at a certain value of $(C,A)$. In Sec.~\ref{subsec:VI.A_interferometric_detection_schemes}
we show that $\sqrt{P_{\{r_{k}=0\}}^{(d)}}e^{i\chi_{\{r_{k}=0\}}^{(d)}}$
is an observable quantity. Therefore, the phase $\chi_{\{r_{k}=0\}}^{(d)}(\theta)$
being ill-defined at $(C_{\mathrm{crit}},A_{\mathrm{crit}})$ implies
$P_{\{r_{k}=0\}}^{(d)}(\theta,C_{\mathrm{crit}},A_{\mathrm{crit}})=0$.
The converse is not necessarily true. However, in our study of measurement-induced
phases we have not found instances of the postselection probability
vanishing without a phase singularity.

In Sec.~\ref{sec:IVB_postsel_crit_line}, we find the set of all
points $(C_{\mathrm{crit}},A_{\mathrm{crit}},\theta_{\mathrm{crit}})$
corresponding to the postselection probability vanishing. Before proceeding
there, we now present a different view of what happens at these special
points.

By construction (cf.~Eqs.~(\ref{eq:def_postselected_d+}--\ref{eq:def_postselected_d-})),
for each given $\theta$, $\chi_{\{r_{k}=0\}}^{(d)}(\theta)$ is defined
modulo $2\pi$. It follows from Eq.~(\ref{eq:postselected_result})
that $\chi_{\{r_{k}=0\}}^{(d)}(\theta=0)=\chi_{\{r_{k}=0\}}^{(d)}(\theta=\pi)=0\,(\mathrm{mod}\,2\pi)$.
Without loss of generality, one can assign $\chi_{\{r_{k}=0\}}^{(d)}(\theta=0)=0$.
On top of that, demanding the continuity of $\chi_{\{r_{k}=0\}}^{(d)}(\theta)$
as a function of $\theta$, one removes the freedom of adding multiples
of $2\pi$ to $\chi_{\{r_{k}=0\}}^{(d)}(\theta)$. One thus must have
$\chi_{\{r_{k}=0\}}^{(d)}(\theta=\pi)=2\pi n$, where $n$ is a well-defined
integer that characterizes the entire dependence of $\chi_{\{r_{k}=0\}}^{(d)}$
on $\theta$ at a given $(C,A)$. It is natural to denote $n$ as
the winding number, as it represents the number of times the function
$e^{i\chi_{\{r_{k}=0\}}^{(d)}(\theta)}$ winds around the origin in
the complex plane. Being an integer number, 
\begin{equation}
n=\frac{1}{2\pi}\int_{0}^{\pi}d\theta\frac{d\chi_{\{r_{k}=0\}}^{(d)}(\theta)}{d\theta}=\frac{\chi_{\{r_{k}=0\}}^{(d)}(\pi)-\chi_{\{r_{k}=0\}}^{(d)}(0)}{2\pi}\label{eq:winding_number_postselected}
\end{equation}
cannot change as $\chi_{\{r_{k}=0\}}^{(d)}(\theta)$ is smoothly deformed,
rendering $n$ a topological invariant. The presence of different
values of $n$ at different measurement parameters $C$ and $A$ implies
existence of a sharp transition where the value of $n$ jumps discontinuously.
In other words, there must exist some critical $(C_{\mathrm{crit}},A_{\mathrm{crit}})$
at which the function $\chi_{\{r_{k}=0\}}^{(d)}(\theta)$ is ill-defined;
it is sufficient for $\chi_{\{r_{k}=0\}}^{(d)}(\theta)$ not to be
well-defined at a single $\theta=\theta_{\mathrm{crit}}$. As discussed
above, this requires $P_{\{r_{k}=0\}}^{(d)}(\theta_{\mathrm{crit}})=0$.
Hence, such transitions between different values of the winding number
$n$ correspond to singularities like the one found above.

Such transitions have been reported in Ref.~\citep{Gebhart2020}
for the case of $A=0$. There, the existence of such transitions is
evident through a simple consideration. For the limit of infinitely
weak measurements ($C=A=0$), $\chi_{\{r_{k}=0\}}^{(d)}(\theta)\equiv0$
yielding $n=0$, while in the limit of projective measurements ($C\rightarrow\infty,A=0$),
$\chi_{\{r_{k}=0\}}^{(d)}(\theta)=\pi d(\cos\theta-1)$ yielding $n=-d$.
Therefore, there must be a transition at some finite $C>0$ when $A=0$.

\begin{figure}
\begin{centering}
\includegraphics[width=1\columnwidth]{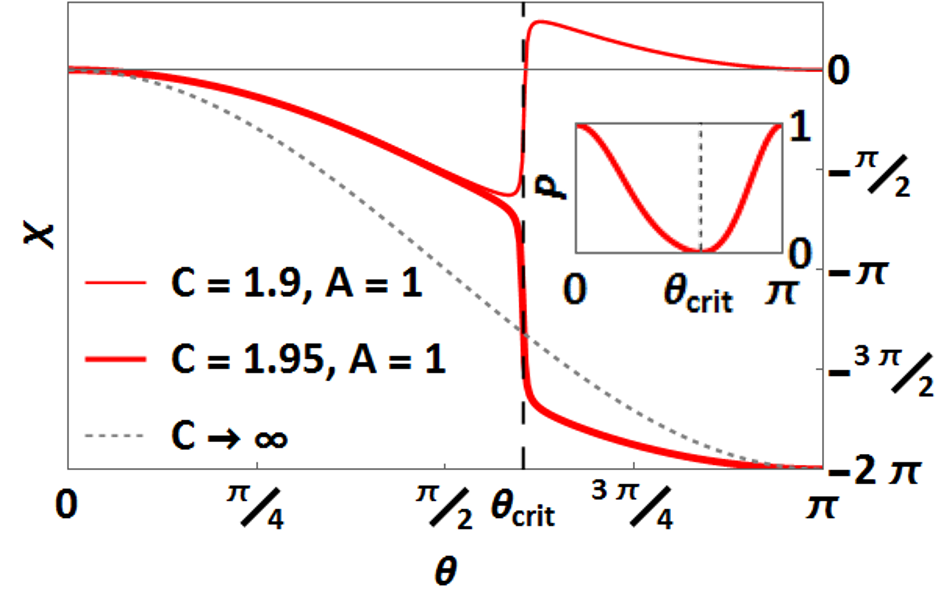}
\par\end{centering}
\caption{\label{fig:transition_postselected_example}The phase $\chi_{\{r_{k}=0\}}^{(d=+1)}\equiv\chi$,
cf.~Eq.~(\ref{eq:postselected_result}), as a function of $\theta$
for $A=1$ for $C$ above and below the critical value $C_{\mathrm{crit}}\approx1.925$.
The winding number $n$, cf.~Eq.~(\ref{eq:winding_number_postselected}),
is equal to $0$ for $C<C_{\mathrm{crit}}$ and to $-1$ for $C>C_{\mathrm{crit}}$.
The behavior of $\chi(\theta<\theta_{\mathrm{crit}})$ immediately
above and below the transition is identical, while the dependence
of $\chi(\theta>\theta_{\mathrm{crit}})$ differs by a $2\pi$ shift.
This leads to $\chi(\theta=\theta_{\mathrm{crit}},C=C_{\mathrm{crit}})$
being ill-defined. The dependence of $P_{\{r_{k}=0\}}^{(d=+1)}\equiv P$
on $\theta$ at $(C=C_{\mathrm{crit}},A=1)$ is shown in the inset.
The undefinedness of $\chi(\theta=\theta_{\mathrm{crit}},C=C_{\mathrm{crit}})$
is enabled by $P(\theta=\theta_{\mathrm{crit}},C=C_{\mathrm{crit}})=0$.}
\end{figure}

For the present, more general case, the above consideration does not
apply. While at $C\rightarrow\infty$, $\chi_{\{r_{k}=0\}}^{(d)}(\theta)=\pi d(\cos\theta-1)$
and $n=-d$ for any $A$, cf.~Eq.~(\ref{eq:postselected_result}),
the phase at $C=0,A\neq0$ is not identically zero. Therefore, one
cannot guarantee the existence of a transition at a certain $C$ for
an arbitrary value of $A$. We find that transitions exist for $\abs A\leq A_{0}=\pi\sqrt{3}/2$
(cf.~Sec.~\ref{sec:IVB_postsel_crit_line} and Appendix~\ref{sec:appendix_critical_line_postselected}),
and do not exist otherwise. An example of such a transition is presented
in Fig.~\ref{fig:transition_postselected_example}.

\begin{figure}
\begin{centering}
\includegraphics[width=1\columnwidth]{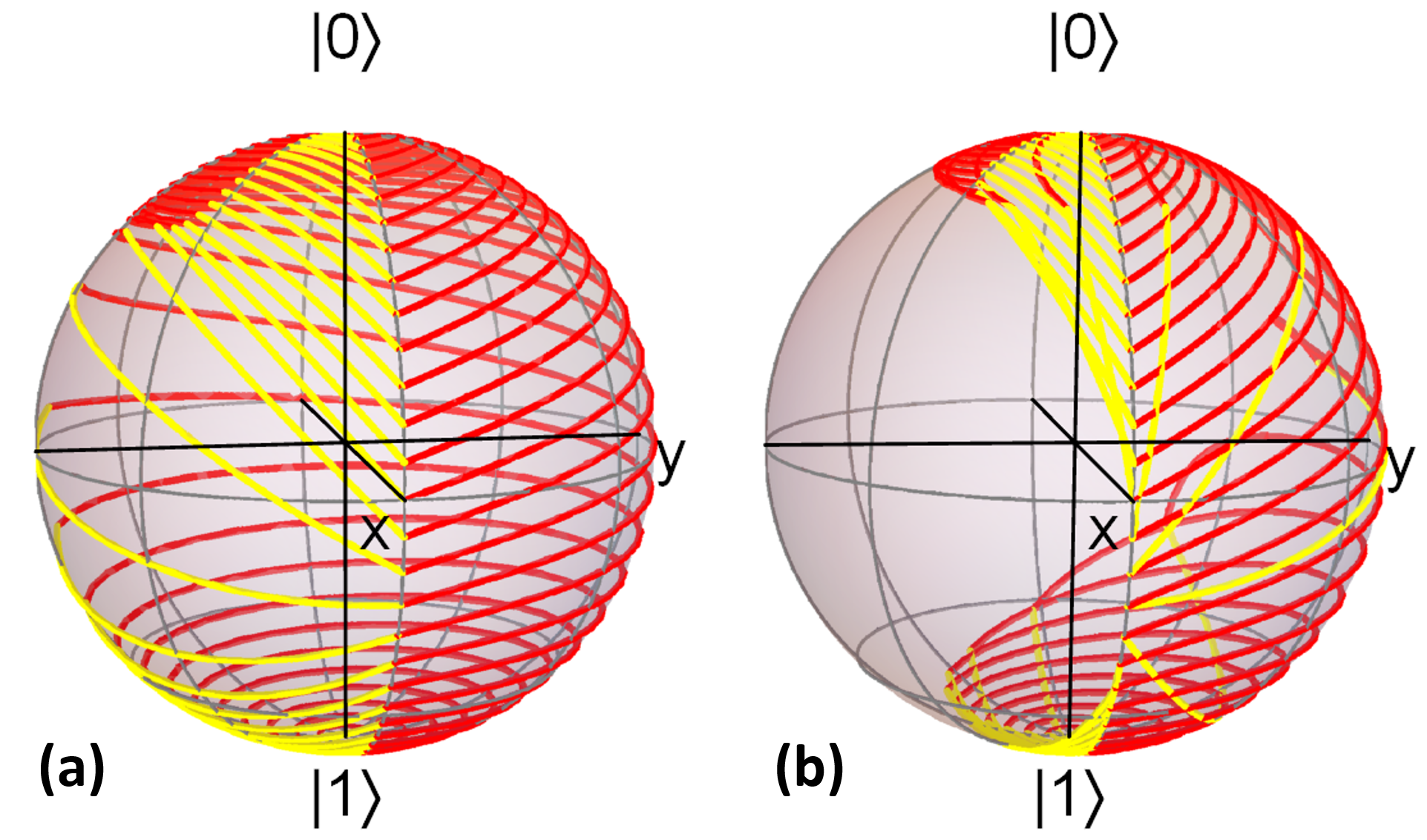}
\par\end{centering}
\caption{\label{fig:trajectories_surface}Measurement-insuced system trajectories
$\{\protect\ket{\psi_{k}}\}$ on the Bloch sphere for $C=2.3>C_{\mathrm{crit}}$
(a) and for $C=1.5<C_{\mathrm{crit}}$ (b), above and below the transition
for $A=1$. Different trajectories correspond to the protocols executed
at different $\theta$. The red segments correspond to the quasicontinuous
sequences $\{\protect\ket{\psi_{k=0,...,N}}\}$, while the yellow
segments are the shortest geodesics on the Bloch sphere connecting
$\protect\ket{\psi_{N}}$ with $\protect\ket{\psi_{N+1}}\propto\protect\ket{\psi_{0}}$.
Above the critical measurement strength, the surface wraps around
the Bloch sphere, while below, it does not. Confer Fig. 5 of Ref.~\citep{Gebhart2020}
for the special case of $A=0$.}
\end{figure}

Reference \citep{Gebhart2020} also linked this type of transitions
to a topological transition of the surface formed by measurement-induced
trajectories on the Bloch sphere. Consider the sequence of states,
$\{\ket{\psi_{k=0,...,N}}\}$, cf.~Eq.~(\ref{eq:def_intermediate_states}),
through which the system passes under the sequence of measurements.
For a quasicontinuous sequence of measurements they form a quasicontinuous
trajectory on the Bloch sphere. This trajectory is not closed. It
can be argued \citep{Chruscinski2004,Gebhart2020} that the natural
way of connecting $\ket{\psi_{N}}$ with $\ket{\psi_{N+1}}\propto\ket{\psi_{0}}$
is by drawing the shortest geodesic on the Bloch sphere, which corresponds
to a postselected projective measurement at the end of the measurement
sequence, cf.~the discussion between Eqs.~(\ref{eq:def_intermediate_states})
and (\ref{eq:chi_definition}). This guarantees that the trajectory
is closed. Consider now all trajectories induced when executing the
protocol at different $\theta\in[0;\pi]$ for a given $(C,A)$. They
form a surface on the Bloch sphere (cf.~Fig.~\ref{fig:trajectories_surface}).
We have found numerically that for $C>C_{\mathrm{crit}}$ the surface
always covers the Bloch sphere, while for $C<C_{\mathrm{crit}}$ it
never does. Therefore, the link between the winding number of the
measurement-induced phase and the topology of the surface formed by
the measurement-induced trajectories exists beyond the case of $A=0$,
studied in Ref.~\citep{Gebhart2020}, notwithstanding the phase not
being immediately related to the trajectory (cf. the discussion in
Sec.~\ref{sec:IIC_dyn/geom+sym/antisym_classifications}).\footnote{We emphasize that the two transitions (in the phase winding number
and in the topology of the surface formed by the measurement-induced
trajectories) always happen concomitantly. In particular, the transition
in the surface topology never takes place at $\abs A>A_{0}=\pi\sqrt{3}/2$.}

\subsection{\label{sec:IVB_postsel_crit_line}The critical line of the transition}

\begin{figure*}
\begin{centering}
\includegraphics[width=1\textwidth]{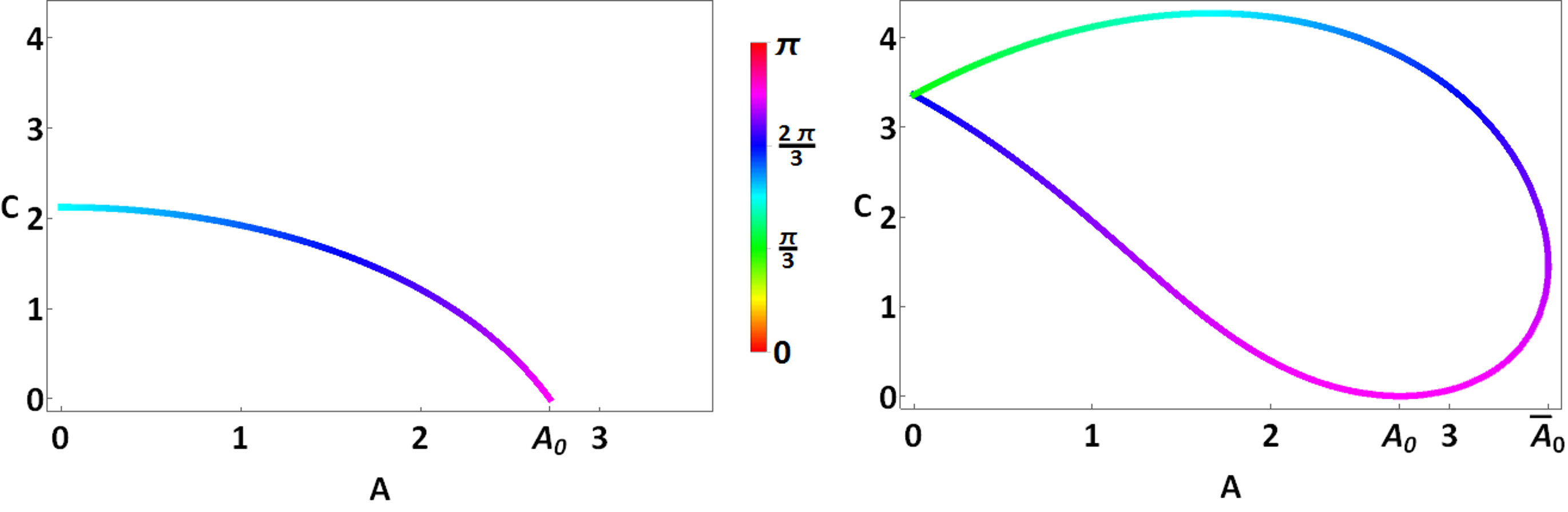}
\par\end{centering}
\caption{\label{fig:critical_line} The critical lines of the topological transitions
in the behavior of the weak-measurement-induced phase for the postselective
($\chi_{\{r_{k}=0\}}^{(d=+1)}$, left panel) and averaging ($\bar{\chi}^{(d=+1)}$,
right panel) protocols. The lines follow the coordinates $(C_{\mathrm{crit}},A_{\mathrm{crit}})$.
The critical polar angle $\theta_{\mathrm{crit}}^{(d=+1)}$ is shown
by the color code. For $d=-1$, the transitions take place at the
same $(C_{\mathrm{crit}},A_{\mathrm{crit}})$ but at $\theta_{\mathrm{crit}}^{(d=-1)}=\pi-\theta_{\mathrm{crit}}^{(d=+1)}$,
as can be inferred from the symmetries discussed in Secs.~\ref{sec:IIIC_postselected_analytics}
and \ref{sec:IIID_averaged_semianalytics}. The behavior at $A<0$
can be inferred too employing those symmetries.}
\end{figure*}

We have demonstrated in Sec.~\ref{sec:IVA_postsel_transition_essence}
that there exist special points $(C_{\mathrm{crit}},A_{\mathrm{crit}},\theta_{\mathrm{crit}}^{(d)})$
where $P_{\{r_{k}=0\}}^{(d)}(\theta_{\mathrm{crit}}^{(d)})=0$. These
special points are associated with phase winding features in the $(C,A)$
plane and with jumps in the winding number $n(C,A)$, cf.~Eq.~(\ref{eq:winding_number_postselected}).
In fact, the set of all these points forms a critical line shown in
Fig.~\ref{fig:critical_line} (left panel). The derivation of this
result is presented in Appendix~\ref{sec:appendix_critical_line_postselected}.

Note that the critical line for the postselective protocol ($\{r_{k}=0\}$)
splits the $(C,A)$ plane into two regions (``phases'', cf.~Fig.~\ref{fig:critical_line}
(left panel)). These correspond to two different values of the winding
number $n$. The region below the critical line corresponds to $n=0$
(a topologically trivial phase). Indeed, at $(C,A)=(0,0)$ the system
is not influenced at all by the measurements leading to $\chi_{\{r_{k}=0\}}^{(d)}(\theta)\equiv0$
and $n=0$. Changing the value of a topological index requires passing
through a critical point $(C_{\mathrm{crit}},A_{\mathrm{crit}})$
such that $P_{\{r_{k}=0\}}^{(d)}(\theta_{\mathrm{crit}}^{(d)})=0$
at some $\theta=\theta_{\mathrm{crit}}^{(d)}$. Since any point within
this region can be accessed from another point by a continuous variation
of parameters without crossing the critical line, it follows that
$n=0$ throughout this region. Similarly, $(C\rightarrow\infty,A=0)$
corresponds to projective measurement and yields the Pancharatnam
phase with $n=-d$. The same connectivity argument implies that this
is the value of $n$ throughout the region above the critical line.

We note that the transition only happens for $A\leq A_{0}=\pi\sqrt{3}/2$.
While this follows from the solution of the problem (cf.~Appendix~\ref{sec:appendix_critical_line_postselected}),
it is instructive to have an intuitive understanding of this fact.
For this consider the case of $C=0$. The back-action of a $r_{k}=0$
measurement (and only $r_{k}=0$ are obtained when $C=0$) is equivalent
to a Hamiltonian evolution for time $\Delta t=1/N$ in a system with
energy gap $\Delta E=-2A$, cf.~Sec.~\ref{sec:IIIB_measurement_sequences+scaling}
and Eq.~(\ref{eq:asymmetry_as_Hamiltonian_evolution}). Then the
total evolution under all the measurements in the $N\rightarrow\infty$
limit is equivalent to a Hamiltonian evolution for time $T=N\Delta t=1$
with a continuously evolving Hamiltonian, followed by a projective
measurement that ensures the return of the system state to $\ket{\psi_{0}}$.
The rate at which the Hamiltonian parameters are varied is of the
order of $\nu=1/T=1$. For $A\rightarrow\infty$, $\Delta ET\sim A\nu^{-1}=A\gg1$,
so that the evolution is adiabatic; the system state follows meticulously
the measurement/Hamiltonian axis and acquires the adiabatic Berry
phase $\chi_{\{r_{k}=0\}}^{(d)}(\theta)=\pi d(\cos\theta-1)$ leading
to a winding number $n=-d$. For $A<\infty$, the evolution is not
adiabatic, implying that the phase will not coincide with the Berry
phase (cf.~Sec.~\ref{subsec:IIIE1_Berry_limit}). At $A=0$, the
evolution is totally non-adiabatic, the system does not have time
to ``sense'' the change in the Hamiltonian axis, and the acquired
phase $\chi_{\{r_{k}=0\}}^{(d)}(\theta)\equiv0$, so that $n=0$.
It is thus clear that there has to be a transition between the two
winding numbers at some value of $A=A_{0}$, which is depicted in
Fig.~\ref{fig:critical_line} (left panel). The required vanishing
of the postselection probability $P_{\{r_{k}=0\}}^{(d)}(\theta_{\mathrm{crit}}^{(d)})$
at $A=A_{0}$ is due to the last projective measurement, implying
that the state to which the system arrives as a result of the non-adiabatic
Hamiltonian evolution is orthogonal to its initial state.\footnote{ In fact, we find that for any $(C_{\mathrm{crit}},A_{\mathrm{crit}},\theta_{\mathrm{crit}}^{(d)})$
the final state after weak-measurement-induced evolution, $\ket{\psi_{N}}$,
is orthogonal to $\ket{\psi_{N+1}}\propto\ket{\psi_{0}}$, while the
probability of observing the sequence of $\left\{ r_{k=1,...,N}=0\right\} $
is $\sp{\psi_{N}}{\psi_{N}}\neq0$.}

The regimes of the Pancharatnam phase ($C\rightarrow\infty$) and
of the Berry phase ($A\rightarrow\infty$) share the same topological
index $n$. It thus comes with little surprise that they can be smoothly
connected, without crossing any critical lines, as follows from Fig.~\ref{fig:critical_line}
(left panel).

In the next section, we analyze topological transitions of the averaged
phase and discuss the qualitative differences from the transitions
discussed above.

\section{\label{sec:V_top_trans_averaged}Topological transitions in the averaging
protocol}

The behavior of the averaged phase $\bar{\chi}^{(d)}$ and dephasing
factor $e^{-\alpha^{(d)}}$ bear numerous similarities to the postselected
phase $\chi_{\{r_{k}=0\}}^{(d)}$ and postselection probability $P_{\{r_{k}=0\}}^{(d)}$.
However, there are important qualitative differences that manifest
themselves in the topological properties of $\bar{\chi}^{(d)}$.

\begin{figure}
\begin{centering}
\includegraphics[width=1\columnwidth]{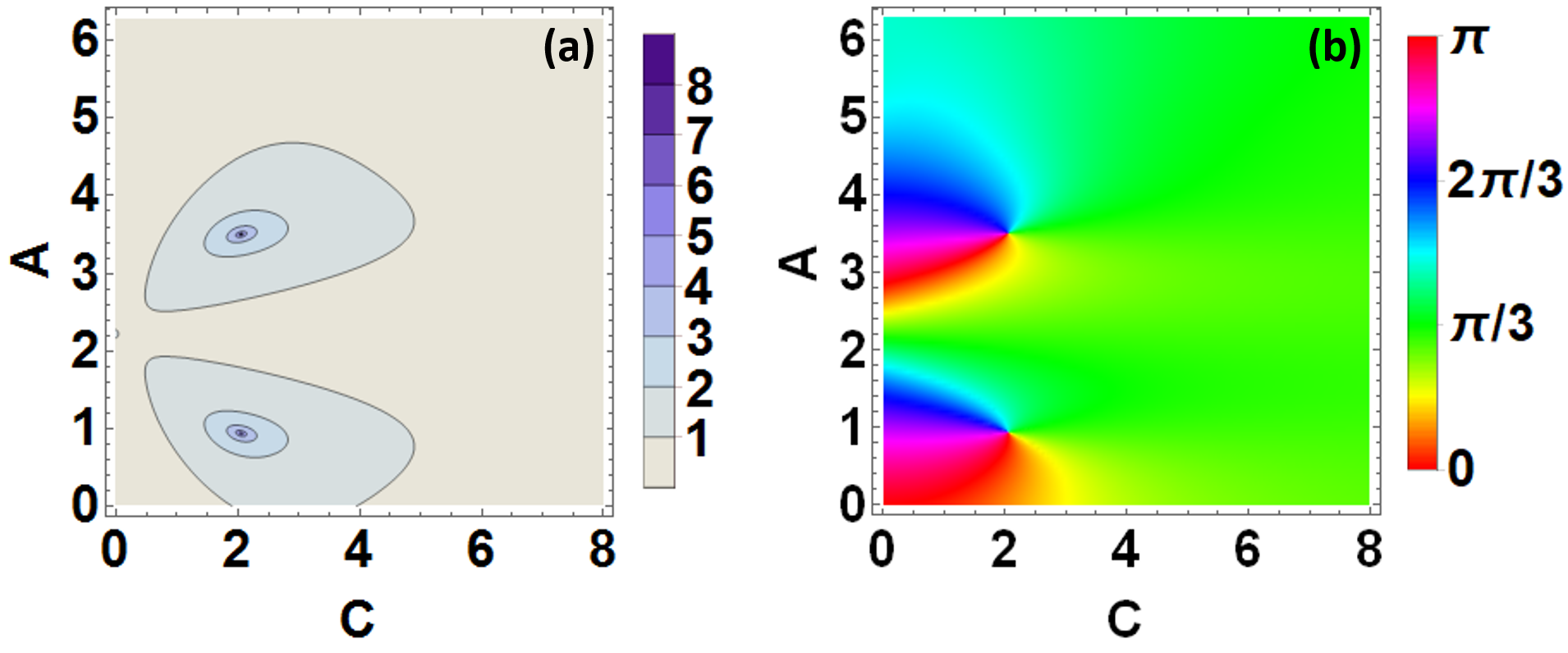}
\par\end{centering}
\caption{\label{fig:avg_dephasing_and_phase}Dephasing $\alpha^{(+1)}$ (a)
and phase $\bar{\chi}^{(+1)}$ (b), cf.~Eq.~(\ref{eq:def_averaged_d}),
at $\theta=3\pi/4$ color-coded as functions of the measurement strength
($C$) and asymmetry ($A$) parameters. Note the two singularities
at $C\approx2$, where $\alpha^{(+1)}$ diverges. The phase makes
$\pi$-windings around the points of divergent $\alpha^{(+1)}$.}
\end{figure}

Similarly to the postselective protocol, the dephasing factor $e^{-\alpha^{(d)}}$
vanishes at specific values $(C_{\mathrm{crit}},A_{\mathrm{crit}},\theta_{\mathrm{crit}}^{(d)})$,
cf.~Fig.~\ref{fig:avg_dephasing_and_phase}(a). Equivalently, one
can say that $\alpha^{(d)}$ diverges at these points. The phase $\bar{\chi}^{(d)}$
makes windings around the points of divergent $\alpha^{(d)}$, cf.~Fig.~\ref{fig:avg_dephasing_and_phase}(b).
However, an important qualitative difference is that $\bar{\chi}^{(d)}$
is defined modulo $\pi$, and not modulo $2\pi$ as the postselected
phases, cf. its definition in Eq.~(\ref{eq:def_averaged_d}). This
implies that the minimum possible winding is of size $\pi$, cf.~Fig.~\ref{fig:avg_dephasing_and_phase}(b),
in contrast to the $2\pi$-windings of $\chi_{\{r_{k}=0\}}^{(d)}$
in Fig.~\ref{fig:logP_divergences_postselected}(b).

The above distinction naturally leads to the fact that the diagram
of topological regimes can be richer in the averaging protocol. Similarly
to the protocol of the postselected phase, we have $e^{2i\bar{\chi}^{(d)}-\alpha^{(d)}}=1$
for $\theta=0$ and $\theta=\pi$, cf.~Eqs.(\ref{eq:MdeltaR_4x4}--\ref{eq:simplification_averaged_largeN}).
However, since the averaged phase $\bar{\chi}^{(d)}$ is defined modulo
$\pi$, and not $2\pi$, the good winding number definition is
\begin{equation}
\bar{n}=\frac{1}{\pi}\int_{0}^{\pi}d\theta\frac{d\bar{\chi}^{(d)}(\theta)}{d\theta}=\frac{\bar{\chi}^{(d)}(\pi)-\bar{\chi}^{(d)}(0)}{\pi}.\label{eq:winding_number_averaged}
\end{equation}
Similarly to $n$ in Eq.~(\ref{eq:winding_number_postselected}),
$\bar{n}$ is also integer-valued but demonstrates a larger spectrum
of values. Indeed, in the limit of $C\rightarrow\infty$ $\bar{\chi}^{(d)}(\theta)=\pi d(\cos\theta-1)$,
implying $\bar{n}=-2d$; for $C=A=0$, $\bar{\chi}^{(d)}(\theta)\equiv0$
and $\bar{n}=0$; yet it is also possible to have $\bar{n}=-d$.

\begin{figure}
\begin{centering}
\includegraphics[width=1\columnwidth]{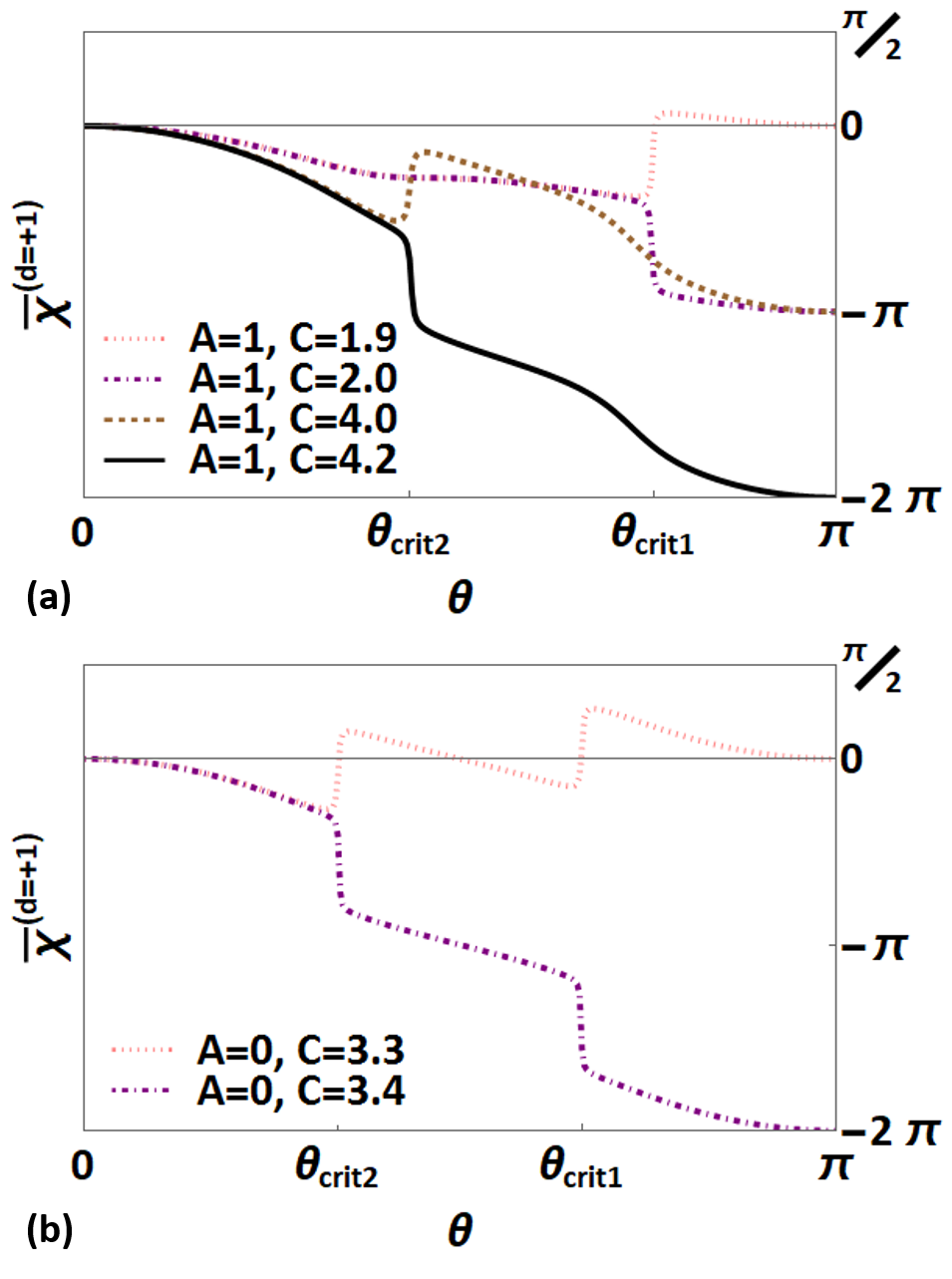}
\par\end{centering}
\caption{\label{fig:transition_averaged_example}The dependence of the averaged
phase $\bar{\chi}^{(d=+1)}(\theta)$ on $\theta$ for various $C$
at $A=1$ (a) and at $A=0$ (b).}
\end{figure}

Switching between different values of $\bar{n}(C,A)$ can only happen
at $(C_{\mathrm{crit}},A_{\mathrm{crit}})$ for which there exists
$\theta_{\mathrm{crit}}^{(d)}$ such that $e^{-\alpha^{(d)}(C_{\mathrm{crit}},A_{\mathrm{crit}},\theta_{\mathrm{crit}}^{(d)})}=0$
(making the phase $\bar{\chi}^{(d)}(C_{\mathrm{crit}},A_{\mathrm{crit}},\theta_{\mathrm{crit}}^{(d)})$
undefined). The set of $(C_{\mathrm{crit}},A_{\mathrm{crit}})$ forms
a critical line (Fig.~\ref{fig:critical_line} (right panel)) separating
the regimes of different $\bar{n}$. Here, the critical line splits
the $(C,A)$ plane into three regions. The outermost and the innermost
regions correspond to $\bar{n}=-2d$ and $\bar{n}=0$ respectively.
The middle one, which was absent in the postselective protocol, corresponds
to $\bar{n}=-d$, cf.~Fig.~\ref{fig:transition_averaged_example}(a).
We emphasize that the $\bar{n}=-d$ region can only be explored with
measurements that have non-Hermitian back-action operators (i.e.,
$A\neq0$).

It is noteworthy that the presence of a middle region is facilitated,
yet not dictated, by the definition of the averaged phase modulo $\pi$.
Indeed, as a matter of principle, one could define the postselected
phase via $P_{\{r_{k}\}}^{(d)}e^{2i\chi_{\{r_{k}\}}^{(d)}}=\bra{\psi_{0}}\mathcal{M}_{N}^{(r_{N})}...\mathcal{M}_{1}^{(r_{1})}\ket{\psi_{0}}^{2}$
as opposed to Eq.~(\ref{eq:chi_definition}). This would imply that
$\chi_{\{r_{k}\}}^{(d)}$ too (not only $\bar{\chi}^{(d)}$) is defined
modulo $\pi$. Nevertheless, the winding properties of the function
$\chi_{\{r_{k}\}}^{(d)}(\theta)$ would not change with the change
of definition. In particular, the winding number $n$ of $\chi_{\{r_{k}=0\}}^{(d)}(\theta)$
would still acquire only two values, $0$ and $-d$ (translating to
$\bar{n}=0$ and $-2d$ respectively). This demonstrates the non-trivial
effect of averaging over the readout sequences.

\begin{figure}
\begin{centering}
\includegraphics[width=1\columnwidth]{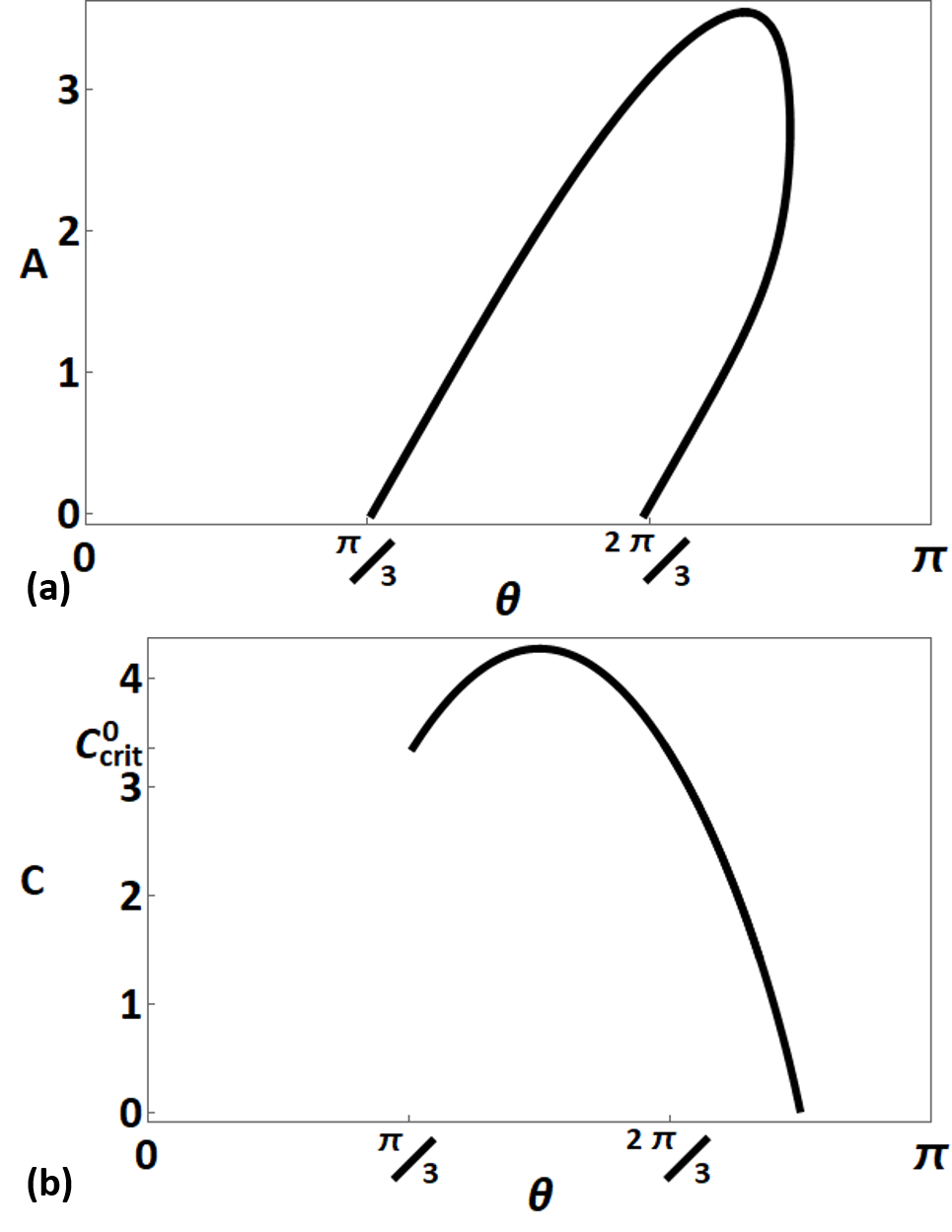}
\par\end{centering}
\caption{\label{fig:AC_vs_theta}Projections of the critical line $(C_{\mathrm{crit}},A_{\mathrm{crit}},\theta_{\mathrm{crit}}^{(d=+1)})$
for the averaged phase, cf.~Fig.~\ref{fig:critical_line} (right
panel), onto the $(A,\theta)$ (a) and $(C,\theta)$ (b) planes.}
\end{figure}

Several other features of the critical line behavior, cf.~Fig.~\ref{fig:critical_line}
(right panel), are noteworthy. First, note that transitions as a function
of $C$ happen only at $A\leq\bar{A}_{0}\approx3.55$. However, the
threshold value is different from that in the postselective protocol:
$A_{0}<\bar{A}_{0}$.

Second, for any $A\in(0;\bar{A}_{0})$ there are \emph{two} transitions,
which correspond to two different critical polar angles, $\theta_{\mathrm{crit}}^{(d)}$.
At $A=0$ there is only one transition at $C=C_{\mathrm{crit}}^{0}\approx3.35$
taking $\bar{n}$ from $0$ to $-2d$. The critical polar angles of
the two transitions do not merge as $A=0$ is approached. This might
be puzzling. The resolution of the puzzle is that at $A=0$, the transition
happens as $\bar{\chi}^{(d)}(\theta)$ exhibits two jumps at different
values of $\theta$, cf.~Fig.~\ref{fig:transition_averaged_example}(b)
and Fig.~\ref{fig:AC_vs_theta}(a).

\begin{figure}
\begin{centering}
\includegraphics[width=1\columnwidth]{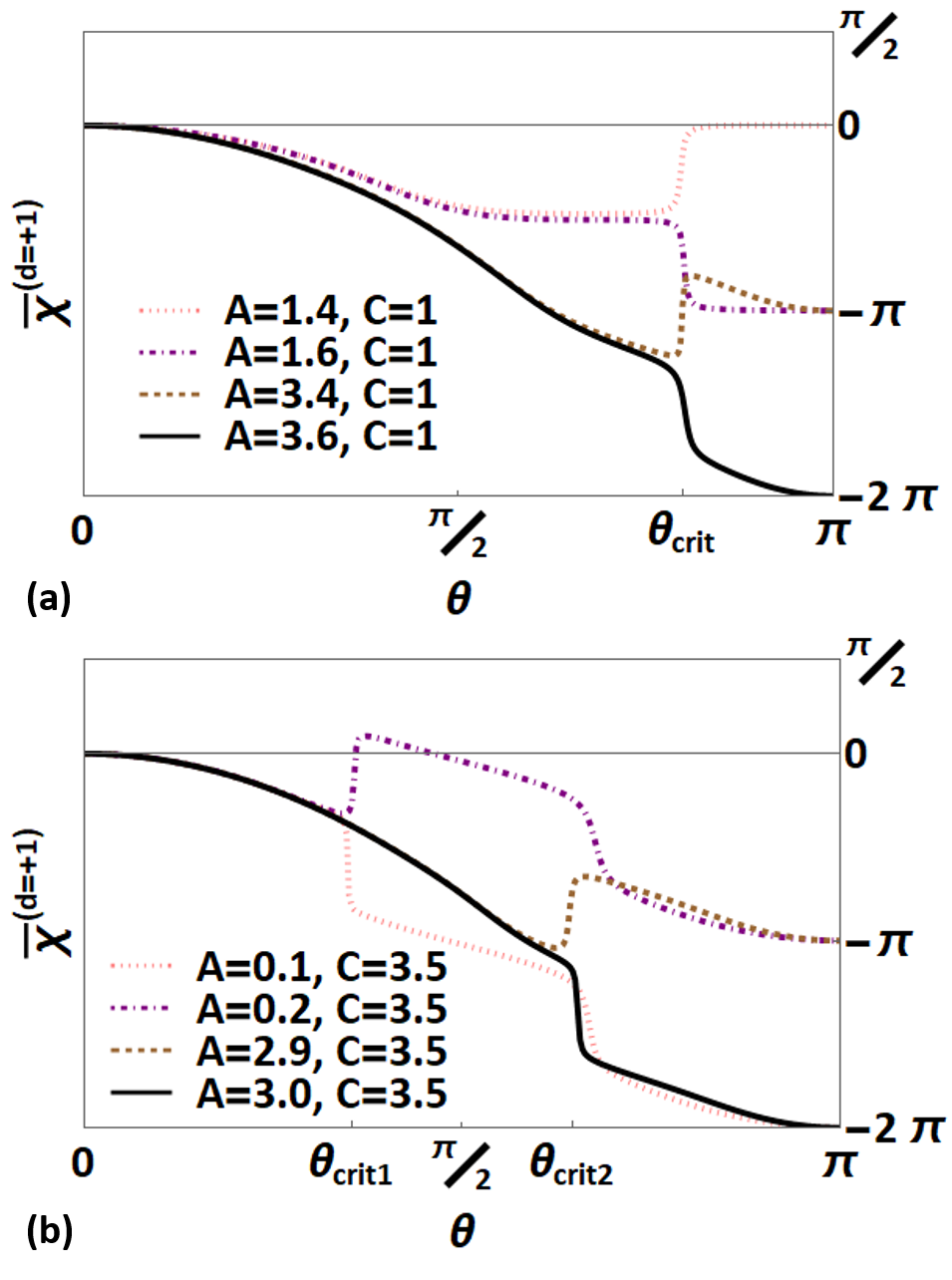}
\par\end{centering}
\caption{\label{fig:transition_averaged_example2}The dependence of the averaged
phase $\bar{\chi}^{(d=+1)}(\theta)$ on $\theta$ for various $A$
at $C=1$ (a) and $C=3.5$ (b).}
\end{figure}

Third, for $C<C_{\mathrm{crit}}^{0}\approx3.35$, there are two transitions
($\bar{n}=0\rightarrow-d$ and $-d\rightarrow-2d$) as a function
of $A$ happening at \emph{the same} value of $\theta$ (cf.~Fig.~\ref{fig:critical_line},
Fig.~\ref{fig:AC_vs_theta}(b), and Fig.~\ref{fig:transition_averaged_example2}(a)).
For $C>C_{\mathrm{crit}}^{0}$, there are again two distinct $A_{\mathrm{crit}}$,
yet now the transitions correspond to $\bar{n}=-2d\rightarrow-d$
and $-d\rightarrow-2d$, and happen at two \emph{different} values
of $\theta$, cf.~Fig.~\ref{fig:transition_averaged_example2}(b).

Finally, both the averaged and the postselected transition have the
same $(A_{\mathrm{crit}},\theta_{\mathrm{crit}}^{(d=+1)})$ at $C=0$.
This is easy to understand, as at $C=0$ essentially no measurement
takes place and there is no difference between $\bar{\chi}^{(d)}(\theta)$
and $\chi_{\{r_{k}\}}^{(d)}(\theta)$: the only readout sequence that
can be obtained is $\{r_{k}=0\}$, cf.~Sec.~\ref{subsec:IIIE3_Hamiltonian_limit}.
However, already at arbitrarily small $C$\textcolor{blue}{,} the
critical lines of the two protocols behave in drastically different
ways.

In Sec.~\ref{sec:IVA_postsel_transition_essence} we linked the topological
transitions in the postselective protocol to a change in the collective
properties of measurement-induced state trajectories, cf.~Fig.~\ref{fig:trajectories_surface}.
Establishing such a connection for the averaging protocol is not straightforward
since averaging over the detector readouts in the phase definition,
cf.~Eq.~(\ref{eq:def_averaged_d}), implies that many measurement-induced
trajectories are involved for each value of protocol parameters. Recently,
it has been predicted theoretically \citep{Lewalle2017} and observed
experimentally \citep{Naghiloo2017} that measurement-induced dynamics
can exhibit one or multiple optimal (``most probable'') quantum
trajectories depending on the system parameters. Further, in the case
of multiple optimal trajectories the system may exhibit chaotic behavior
\citep{Lewalle2018}. It would be interesting to investigate whether
transitions between these different regimes exist in our system and,
if they do, whether they are linked to the topological transitions
we report here.

\section{\label{sec:VI_experimental_implementation}Experimental implementation}

In this section, we discuss conceptual experimental setups that enable
observing the measurement-induced phases defined and investigated
above. We pay particular attention to some practical aspects of measuring
the averaged phase.

\subsection{\label{subsec:VI.A_interferometric_detection_schemes}Interferometric
detection schemes}

\begin{figure}
\begin{centering}
\includegraphics[width=1\columnwidth]{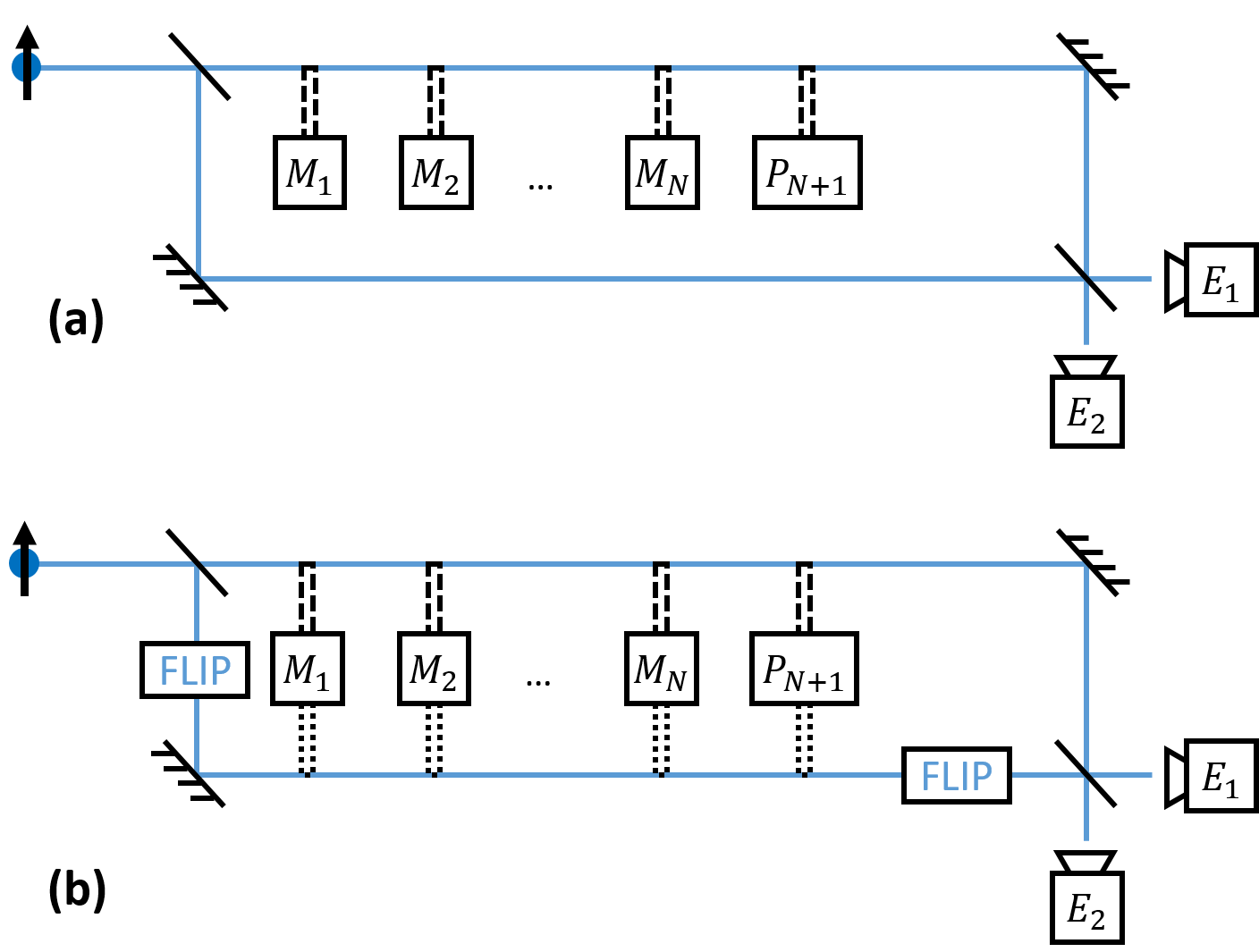}
\par\end{centering}
\caption{\label{fig:interferometers}Interferometry setups for observing measurement-induced
phases $\chi_{\{r_{k}=0\}}$ (a) and $\bar{\chi}$ (b), cf.~Eqs.~(\ref{eq:chi_definition}--\ref{eq:chi-bar_definition}).
A particle, whose spin represents the measured system, flies through
the interferometer. Weak measurements of its spin are denoted by $M_{k=1,...,N}$,
while the last postselected projective measurement is denoted by $P_{N+1}$.
The protocol of Fig.~\ref{fig:interferometers}(b) for detecting
the averaged phase involves two special features. First, the particle
spin is flipped in one arm as indicated by the \textquotedblleft FLIP\textquotedblright{}
boxes. Second, the detectors interact with the two arms via different
Hamiltonians (see Appendix~\ref{sec:appendix_averaged_phase_detection}
for details).}
\end{figure}

In order to measure the effects discussed in the previous sections,
it is crucial to have the ability to access the measurement-induced
phases. Here we define two conceptual setups that facilitate measurement
of the postselected $\chi_{\{r_{k}=0\}}$ and the averaged $\bar{\chi}$
phases (defined in Eqs.~(\ref{eq:chi_definition}) and (\ref{eq:chi-bar_definition})
respectively). The setup for measuring $\chi_{\{r_{k}=0\}}$ is shown
in Fig.~\ref{fig:interferometers}(a). A particle with spin in state
$\ket{\psi_{0}}$ enters a Mach-Zehnder interferometer and is split
into two arms. In one arm, the particle is subjected to a sequence
of weak measurements and one projective measurement (implementing
the protocol described in Sec.~\ref{sec:III_particular_protocol}).
In the other arm, the particle flies through unaffected. As a result,
the state of the particle and detectors just before the particle reaches
the final beam splitter is
\begin{multline}
\ket{\Psi}=\frac{1}{\sqrt{2}}\ket{\psi_{0}}\ket{-1}_{a}\prod_{k=1}^{N+1}\ket{r_{k}=0}_{D_{k}}\\
+\frac{1}{\sqrt{2}}\sum_{\{r_{k}\}}\delta_{r_{N+1},0}\ket{\psi_{0}}\ket{+1}_{a}\\
\times\bra{\psi_{0}}\mathcal{M}_{N}^{(r_{N})}...\mathcal{M}_{1}^{(r_{1})}\ket{\psi_{0}}\prod_{k=1}^{N+1}\ket{r_{k}}_{D_{k}},
\end{multline}
where $\ket{\pm1}_{a}$ denotes the particle being in the upper/lower
arm, and $\ket{r_{k}}_{D_{k}}$ is the state of the $k$th detector.
We have accounted here for the fact that as the particle is flying
through the lower arm, the detectors remain in their initial states
$\ket{r_{k}=0}$ (which are the initial states of the detectors in
the measurement model described in Sec.~\ref{sec:IIIA_detector_model}).
As a result, the intensities observed at the interferometer exits,
$E_{1,2}$, will be
\begin{multline}
I_{1,2}=\frac{I_{0}}{2}\Biggl(\frac{1}{2}+\frac{1}{2}\sum_{\{r_{k}\}}\abs{\bra{\psi_{0}}\mathcal{M}_{N}^{(r_{N})}\dots\mathcal{M}_{1}^{(r_{1})}\ket{\psi_{0}}}^{2}\\
\pm\mathrm{Re}\thinspace\bra{\psi_{0}}\mathcal{M}_{N}^{(0)}...\mathcal{M}_{1}^{(0)}\ket{\psi_{0}}\Biggr),\label{eq:postselected_intensity}
\end{multline}
where $I_{0}$ is the intensity of the incoming particle beam; the
second term on the r.h.s. of Eq.~(\ref{eq:postselected_intensity})
is less than $1/2$ as it accounts for the loss of particles due to
discarding the runs in which the last projective measurement yields
$r_{N+1}=1$; the last---interference---term gives $\sqrt{P_{\{r_{k}=0\}}}e^{i\chi_{\{r_{k}=0\}}}$.
This scheme thus enables the observation of the measurement-induced
phase for the readout sequence $\{r_{k}=0\}$. The scheme relies crucially
on the fact that the readouts $r_{k}=0$ correspond to the detector
initial state being unchanged.

The setup of Fig.~\ref{fig:interferometers}(b) shows how the averaged
phase $\bar{\chi}$ can be measured. Now the particle interacts with
the detectors in both arms. Moreover, the $k$th measurement is performed
in both arms by the same physical detector that is later read out,
thus ensuring that the readout $r_{k}$ is the same in both arms.
However, measuring $e^{2i\bar{\chi}}$ as defined in Eq.~(\ref{eq:chi-bar_definition})
through interference requires that for each readout sequence $\{r_{k}\}$
the particle acquires phase $e^{i\chi_{\{r_{k}\}}}$ in one arm and
$e^{-i\chi_{\{r_{k}\}}}$ in the other arm. In order to achieve that,
we propose to flip the particle spin when it enters and exits the
lower arm and, in addition, to use somewhat different particle-detector
interaction Hamiltonians in the two arms. We give the details of the
procedure in Appendix~\ref{sec:appendix_averaged_phase_detection}.
The resulting intensities at the interferometer exits $E_{1,2}$ are
\begin{multline}
I_{1,2}=\frac{I_{0}}{2}\Biggl(\sum_{\{r_{k}\}}\abs{\bra{\psi_{0}}\mathcal{M}_{N}^{(r_{N})}\dots\mathcal{M}_{1}^{(r_{1})}\ket{\psi_{0}}}^{2}\\
\pm\mathrm{Re}\sum_{\{r_{k}\}}\left(\bra{\psi_{0}}\mathcal{M}_{N}^{(r_{N})}\dots\mathcal{M}_{1}^{(r_{1})}\ket{\psi_{0}}\right)^{2}\Biggr),\label{eq:averaged_intensity}
\end{multline}
where the first term accounts for the particle loss in the last projective
measurement postselection, and the interference term is exactly $e^{2i\bar{\chi}-\alpha}$
in Eq.~(\ref{eq:chi-bar_definition}).\footnote{Equation~(\ref{eq:averaged_intensity}) is valid for arbitrary $N$
for the protocol defined in Sec.~\ref{sec:III_particular_protocol}.
However, Eq.~(\ref{eq:averaged_intensity}) does not apply to protocols
with other choices of the measurement axes $\mathbf{n}_{k}^{(s)}$
and/or the intitial state $\ket{\psi_{0}}$.}

We stress that while for the reasons of theoretical simplification
we have considered the limit of the number of measurements $N\rightarrow\infty$
in the above sections, essentially the same physics of asymmetric
behavior of the phases, the postselection probability, and the dephasing
parameter will appear for sequences of measurements with any $N\geq2$.
Furthermore, the points of vanishing postselection probability and
the singularities of the dephasing parameter will be related to the
topological transitions in the phase behavior for finite $N$ too.
However, some specific features (such as the shape of the critical
lines) will be modified in the case of finite $N$. In particular,
the results will depend periodically on $A$ with the period being
$\pi N$.

\subsection{\label{subsec:VI.B_MC_vs_semianalytics}Some remarks concerning practicalities
of averaging over readout sequences in experiment}

\begin{figure}
\begin{centering}
\includegraphics[width=1\columnwidth]{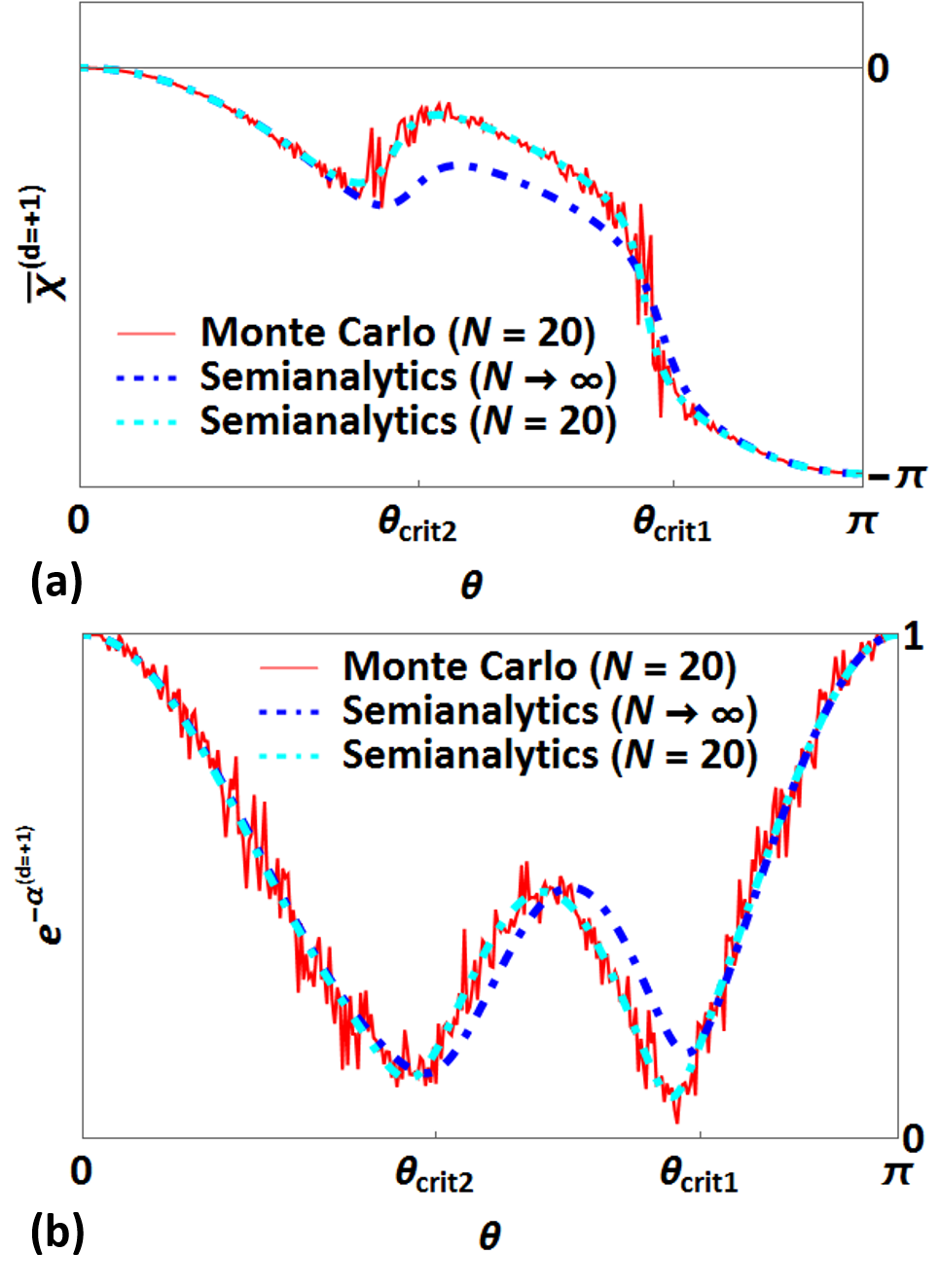}
\par\end{centering}
\caption{\label{fig:monte-carlo-vs-semianalytics}Comparison of Monte Carlo
simulations for the averaged phase $\bar{\chi}^{(d=+1)}$ (a) and
the dephasing parameter $\alpha^{(d=+1)}$ (b) with the results obtained
using the semianalytical method of Sec.~\ref{sec:IIID_averaged_semianalytics}.
The number of Monte Carlo samples (readout sequence realizations $\{r_{k}^{(i)}\}$)
used is $N_{\mathrm{rs}}=100$. The number of measurements in the
sequence is denoted by $N$. The plots correspond to $C=3$, $A=1$.}
\end{figure}

The definition of the averaged phase $\bar{\chi}$ in Eq.~(\ref{eq:chi-bar_definition})
requires averaging the postselected phases $\chi_{\{r_{k}\}}$ over
all possible readout sequences $\{r_{k}\}$ weighted with appropriate
probabilities. However, this does not correspond to the interferometric
procedure for measuring $\bar{\chi}$ outlined above. Indeed, a particle
flying through the interferometer will yield a specific readout sequence
$\{r_{k}\}$ and a specific phase $\chi_{\{r_{k}\}}$ with probability
$P_{\{r_{k}\}}$. The next particle will again yield a random readout
sequence $\{r_{k}'\}$, and so on. Therefore, the actual measurement
procedure is identical to a Monte Carlo sampling of the readout sequences
rather than systematic summing over them. The number of such sequences
scales as $2^{N}$ with the number of measurements $N$. Sampling
such a large number of sequences (for large $N$) is impossible. However,
the probability of a specific sequence $\{r_{k}\}$ determines both
the frequency of obtaining this sequence and its contribution to the
sum, rendering it possible to obtain an accurate estimate of $\bar{\chi}$
with a moderate number of experimental runs.

We have performed a Monte Carlo study that simulates the sampling
of $\{r_{k}\}$ in the experiment. Namely, we randomly generated the
readout sequences according to the algorithm outlined in Ref.~\citep{Gebhart2020}
and calculated 
\begin{equation}
\langle e^{2i\chi_{\{r_{k}\}}}\rangle=\frac{1}{N_{\mathrm{rs}}}\sum_{i=1}^{N_{\mathrm{rs}}}e^{2i\chi_{\{r_{k}^{(i)}\}}},
\end{equation}
where $N_{\mathrm{rs}}=100$ readout sequences $\{r_{k}^{(i)}\}$
were generated for sequences of $N=20$ measurements. A comparison
of the Monte Carlo simulations to the results obtained using the method
of Sec.~\ref{sec:IIID_averaged_semianalytics} is shown in Fig.~\ref{fig:monte-carlo-vs-semianalytics}.
The Monte Carlo curves reproduce the behavior for $N\rightarrow\infty$
qualitatively, and closely follow the exact result for $N=20$. We,
therefore, conclude that the experimental procedure does allow one
to probe the physics discussed above with reasonable accuracy. Although
pinpointing the exact locations of the critical lines of the topological
transitions (where the terms in the sum in Eq.~(\ref{eq:chi-bar_definition})
accurately cancel out to yield $e^{2i\bar{\chi}-\alpha}=0$) may require
a large number of experimental runs, establishing the existence of
several topological sectors with different winding numbers $\bar{n}$
can be done without accumulating too large a statistics.

\section{\label{sec:VII_Conclusions}Conclusions}

We have performed a detailed investigation of measurement-induced
phase factors. Our theory brings forward two classifications of such
phases: dynamical vs. geometrical phases, and components which are
symmetric/antisymmetric with respect to the reversal of the measurement
sequence. Importantly, we have shown based on general considerations
and on a specific example that these two classifications do not coincide.

We have demonstrated our theoretical framework via analyzing a specific
protocol, calculating postselected and averaged measurement-induced
phases, and investigating their dependence on various measurement
parameters. We have shown that the projective-measurement-induced
Pancharatnam phase and the Berry phase induced by adiabatic Hamiltonian
evolution can be viewed as two (out of several) limiting cases of
the phases induced by quasicontinuous sequences of weak measurements.

We have found and investigated topological transitions pertaining
to measurement-induced phases. We have found the ``phase diagram''
of different topological regimes and discussed its distinctive features.
While we have investigated topological transitions for a specific
protocol, the generality of our considerations leads one to believe
that such transitions are a generic feature of measurement-induced
phases, avoiding the need to refer to a specific measurement model
or phase-inducing protocol. Nevertheless, the details of the ``phase
diagram'' may depend on the specific protocol and measurement class.

Finally, we have proposed experimental setups facilitating the observation
of weak-measurement-induced phases and the study of the effects discussed
in this work. We believe that weak-measurement-induced phase factors
present a rich playground that may be important for understanding
topological phases of matter in open quantum systems.

\begin{acknowledgments}

We thank V. Gebhart for useful discussions. We acknowledge funding
by the Deutsche Forschungsgemeinschaft (DFG, German Research Foundation)
-- Projektnummer 277101999 -- TRR 183 (project C01) and Projektnummern
EG 96/13-1, GO 1405/6-1, and MI 658/10-2, and by the Israel Science
Foundation (ISF).

\end{acknowledgments}

\appendix

\section{\label{sec:appendix_scaling_regimes}Investigation of different scaling
regimes}

As mentioned in Sec.~\ref{sec:IIIB_measurement_sequences+scaling},
taking the limit $N\rightarrow\infty$ in our protocols requires adjusting
(scaling) the measurement parameters $g$ and $\theta^{(D)}$, cf.~Sec.~\ref{sec:IIIA_detector_model},
performing this as a function of $N$. Here we explore the possible
ways of scaling. We show that the only non-trivial scaling regime
corresponds to the one presented in Sec.~\ref{sec:IIIB_measurement_sequences+scaling}.

\begin{figure*}
\begin{centering}
\includegraphics[width=1\textwidth]{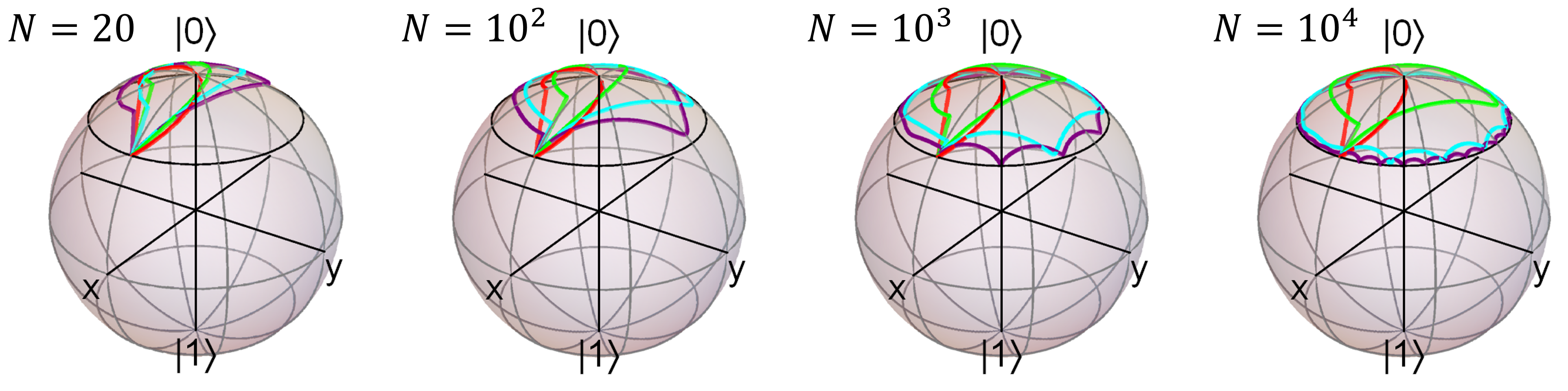}
\par\end{centering}
\caption{\label{fig:scaling-limit-trajectories}Trajectories of the quantum
state, $\protect\ket{\psi_{k}}$ (\ref{eq:def_intermediate_states}),
on the Bloch sphere in various scaling limits. Plotted are the trajectories
induced by sequences of $N$ measurements around the parallel corresponding
to $\theta=\pi/4$ (black) with all measurements yielding readouts
$r_{k}=0$. The back-action matrix $M^{(0)}$ is given in Eq.~(\ref{eq:M0_split_scaling}).
The trajectories are plotted for $b=0$ (purple), $0.1$ (cyan), $0.3$
(green), and $0.5$ (red) using $C=1$, $A=1$, and $d=+1$. All the
trajecotries with $b<1/2$ converge towards the parallel line (i.e.,
the trajectory of the measurement axes) as $N\rightarrow\infty$.
This does not happen for $b=1/2$. Moreover, the trajectory seems
not to change with increasing $N$, suggesting that already at $N=20$
measurements the measurement-induced trajectory has converged.}
\end{figure*}

One can understand the need for scaling of the measurement parameters
in the quasicontinuous limit ($N\rightarrow\infty$) from the following
consideration. Arrange the measurements into $j$ sets ($N\gg j\gg1$).
That is, measurements $k=1,...,N/j$ form one set, $k=N/j+1,...,2N/j$
are the second set etc. Within each set, the axes of the measured
system observables are clustered at $\varphi_{k}\approx\tilde{\varphi}_{k}=2\pi d\left\lfloor (k-1)j/N\right\rfloor /j$,
where $\left\lfloor (k-1)j/N\right\rfloor $ is the set number; here
$\left\lfloor x\right\rfloor $ is the floor function. The spread
of the actual $\varphi_{k}$ from $\tilde{\varphi}_{k}$ is $O(2\pi d/j)$
and can be made arbitrarily small in the limit $N\rightarrow\infty$
via taking arbitrarily large values of $j$. Then this set of weak
measurements can be interpreted as a single projective measurement
with the appropriate axis $\tilde{\mathrm{\mathbf{n}}}_{k}^{(s)}\approx\mathbf{n}_{k}^{(s)}$.
Indeed, if all measurements in a set yield $r_{k}=0$, the back-action
on the system state can be described by $\prod_{k}R^{-1}(\mathbf{n}_{k}^{(s)})M^{(0)}R(\mathbf{n}_{k}^{(s)})\approx R^{-1}(\tilde{\mathbf{n}}_{k}^{(s)})\left(M^{(0)}\right)^{N/j}R(\tilde{\mathbf{n}}_{k}^{(s)})$
with $M^{(0)}$ and $R(\mathbf{n}^{(s)})$ defined in Eqs.~(\ref{eq:back_action_0})
and (\ref{eq:rotation}) respectively. Therefore, $\left(M^{(0)}\right)^{N/j}$
plays the role of the effective back-action matrix $\tilde{M}^{(0)}$.
In the limit $N/j\rightarrow\infty$, with fixed measurement parameters
$g$, $\theta^{(D)}$, $\varphi^{(D)}=-\pi/2$,
\begin{equation}
\tilde{M}^{(0)}=\left(M^{(0)}\right)^{N/j}=\begin{pmatrix}1 & 0\\
0 & 0
\end{pmatrix}
\end{equation}
unless $\sin g\sin\theta^{(D)}=0$. Therefore, for generic measurement
parameters, a set of measurements all yielding $r=0$ is equivalent
to a single projective measurement yielding $r=0$. If at least one
measurement in the set yields $r=1$, the form of $M^{(1)}$ ensures
that the system state is projected onto the $\downarrow$ eigenstate
of $\tilde{\mathrm{\mathbf{n}}}_{k}^{(s)}\cdot\bm{\sigma}^{(s)}$,
again making the back-action identical to that of a single projective
measurement (up to errors $O(2\pi d/j)$). Since the back-action also
determines the probabilities of the outcomes, cf.~Sec.~\ref{sec:IIA_measurement_theory},
one concludes that the original protocol is equivalent to a quasicontinuous
sequence of strong measurements. The latter yields the Pancharatnam
phase.

A non-trivial scaling limit thus requires $\lim_{N/j\rightarrow\infty}\abs{\cos g+i\sin g\cos\theta^{(D)}}^{N/j}>0$.
Since the scaling of the measurement parameters should not depend
on the number of groups $j$ but only on $N$, this is equivalent
to $\lim_{N\rightarrow\infty}\abs{\cos g+i\sin g\cos\theta^{(D)}}^{N}>0$.
Similarly, we require $\lim_{N\rightarrow\infty}\abs{\cos g+i\sin g\cos\theta^{(D)}}^{N}<1$
as the opposite would imply that there is zero probability of obtaining
a $r=1$ readout, making the evolution completely deterministic (equivalent
to Hamiltonian evolution). We allow the following scaling $g=C'N^{-a}$,
$\theta^{(D)}=\pi/2-A'N^{-b}$ with $a,b\geq0$. The above requirements
then imply $a=1/2$. With this choice, the single measurement $r=0$
back-action matrix becomes
\begin{multline}
M^{(0)}=\begin{pmatrix}1 & 0\\
0 & 1-\frac{C'^{2}}{2N}+i\frac{C'A'}{N^{1/2+b}}+O(N^{-3b},N^{-b-3/2})
\end{pmatrix}\\
=\begin{pmatrix}1 & 0\\
0 & \exp\left(-2\frac{C+iAN^{1/2-b}}{N}+O(N^{-3b},N^{-b-3/2})\right)
\end{pmatrix},
\end{multline}
where in the last step we defined $C=C'^{2}/4$ and $A=-A'C'/2$.

We now need to choose the appropriate scaling of $b$. Again, a qualitative
consideration is useful here. Note that
\begin{multline}
M^{(0)}=\begin{pmatrix}1 & 0\\
0 & \exp\left(-2\frac{C+iAN^{1/2-b}}{N}\right)
\end{pmatrix}\\
=\begin{pmatrix}1 & 0\\
0 & \exp\left(-2\frac{C}{N}\right)
\end{pmatrix}e^{-iH\Delta t},\label{eq:M0_split_scaling}
\end{multline}
where $H=B\left(\mathbb{I}-\sigma_{z}^{(s)}\right)$ with $B=AN^{1/2-b}$
and $\Delta t=1/N$. In other words, a measurement of the class we
consider, when yielding $r=0$, can be decomposed into a Hamiltonian
evolution over time $\Delta t$, followed by a measurement with a
Hermitian back-action matrix. Let us first ignore the measurement
(put $C=0$). Then the readout $r=0$ is implied since $M^{(1)}=0$,
cf.~Eq.~(\ref{eq:back_action_1}). The evolution is a quasicontinuous
Hamiltonian evolution, with the magnetic field axis changing its direction
by $\Delta\varphi=2\pi d\Delta t$ after every time interval $\Delta t$.
In the limit $\Delta t\rightarrow0$ (or equivalently, $N\rightarrow\infty$),
this becomes a continuously evolving Hamiltonian with its axis changing
at the rate $d\varphi/dt=2\pi d$, and the energy gap $2AN^{1/2-b}$.
Therefore, for $b<1/2$, the adiabatic theorem applies to the system
as the gap size is infinitely large; the system state will then meticulously
follow the measurement axis. Now, introducing $C>0$ does not modify
the picture much as the corresponding part of the back-action pulls
the state closer towards the measurement axis, to which the state
is close anyway. Numerical investigation shows that this qualitative
consideration is correct, cf.~Fig.~\ref{fig:scaling-limit-trajectories}.
Moreover, one can analytically show that for $b<1/2$, the probability
of getting all measurement outcomes $r_{k}=0$, $P_{\{r_{k}=0\}}^{(d)}=1$,
while the phase $\chi_{\{r_{k}=0\}}^{(d)}=-\pi d(1-\cos\theta)$ coincides
with the Berry phase for the corresponding Hamiltonian evolution.
The regime of $b>1/2$ is also not interesting for us as the non-Hermitian
part becomes insignificant when $N\rightarrow\infty$ and the problem
reduces to that investigated in Ref.~\citep{Gebhart2020}.

We conclude that the non-trivial scaling regime corresponds to $a=b=1/2$
with $g=\sqrt{4C/N}$ and $\theta^{(D)}=\pi/2+A/\sqrt{CN}$. This
is the regime presented in Sec.~\ref{sec:IIIB_measurement_sequences+scaling}.

\section{\label{sec:appendix_critical_line_postselected}Finding the critical
line for the postselective protocol}

\begin{figure*}
\begin{centering}
\includegraphics[width=1\textwidth]{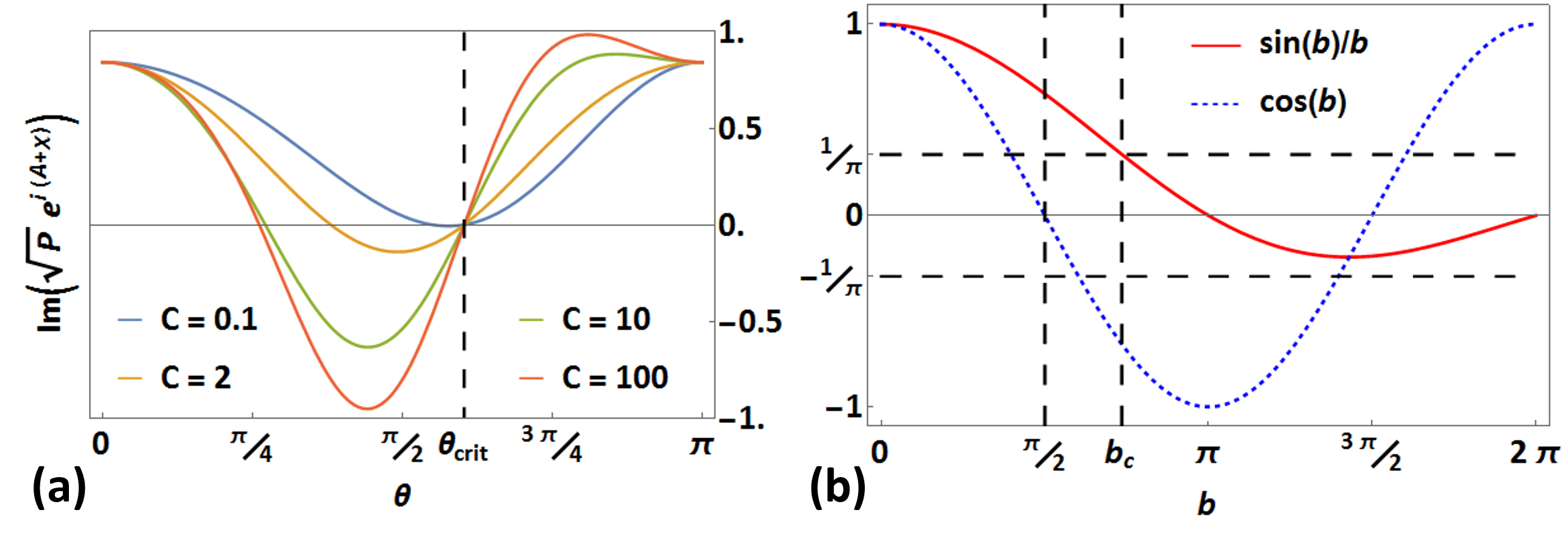}
\par\end{centering}
\caption{\label{fig:imaginary-part-crossing_PS}(a) The imaginary part $\mathrm{Im}\left[\sqrt{P_{\{r_{k}=0\}}^{(d=+1)}}e^{i\chi_{\{r_{k}=0\}}^{(d=+1)}+iA}\right]\equiv\mathrm{Im}\left[\sqrt{P}e^{i(A+\chi)}\right]$
as a function of $\theta$ for various $C$ at $A=1$. One can check
that the probability $P_{\{r_{k}=0\}}^{(d=+1)}=0$ at $(C_{\mathrm{crit}}\approx1.925,A_{\mathrm{crit}}=1,\theta_{\mathrm{crit}}\approx1.894)$.
The imaginary part vanishes at $\theta=\theta_{\mathrm{crit}}$ for
any $C$. (b) The dependence of $\sin b/b$ and $\cos b$ on $b$.
$b_{c}$ is such that $\sin b_{c}/b_{c}=1/\pi$.}
\end{figure*}

Here, we look for the parameters $(C_{\mathrm{crit}},A_{\mathrm{crit}},\theta_{\mathrm{crit}}^{(d)})$
where $P_{\{r_{k}=0\}}^{(d)}=0$. We focus on $A\geq0$. The critical
line at $A<0$ can be inferred using the symmetries discussed at the
end of Sec.~\ref{sec:IIIC_postselected_analytics}. Equation (\ref{eq:postselected_result})
implies that $P_{\{r_{k}=0\}}^{(d)}=0$ is equivalent to 
\begin{equation}
\cosh\tau+Z\frac{\sinh\tau}{\tau}=0.\label{eq:cosh+...=00003D0}
\end{equation}
The real and imaginary parts of this equation represent two equations
for three parameters, $C$, $A$, and $\theta$. Therefore, one expects
the solution to represent a line in the space of these three parameters.

We first observe numerically that if Eq.~(\ref{eq:cosh+...=00003D0})
holds at some $(C_{\mathrm{crit}},A_{\mathrm{crit}},\theta_{\mathrm{crit}}^{(d)})$,
then the imaginary part of the l.h.s. of Eq.~(\ref{eq:cosh+...=00003D0})
vanishes at $(C,A_{\mathrm{crit}},\theta_{\mathrm{crit}}^{(d)})$
for any $C$, cf.~Fig.~\ref{fig:imaginary-part-crossing_PS}(a).
Assuming that this observation is exact, we extract the relation between
$A_{\mathrm{crit}}$ and $\theta_{\mathrm{crit}}^{(d)}$ as follows.
We expand Eq.~(\ref{eq:cosh+...=00003D0}) at $C\rightarrow\infty$
and demand that its imaginary part vanishes to the leading order in
$C$, obtaining the condition
\begin{equation}
A_{\mathrm{crit}}+\pi d\cos\theta_{\mathrm{crit}}^{(d)}=0.\label{eq:crit_A_theta_postselected}
\end{equation}
It is now straightforward to verify that this condition indeed implies
that the imaginary part of the l.h.s. of Eq.~(\ref{eq:cosh+...=00003D0})
vanishes. Indeed, when Eq.~(\ref{eq:crit_A_theta_postselected})
holds, $Z=C$, and $\tau=\sqrt{Z^{2}-\pi^{2}\sin^{2}\theta_{\mathrm{crit}}^{(d)}}$
is either purely real or purely imaginary; therefore, $Z$, $\cosh\tau$,
$\sinh\tau/\tau$ are all real.

Switching to the real part of Eq.~(\ref{eq:cosh+...=00003D0}), we
rewrite the equation as
\begin{equation}
\tau\coth\tau=-Z.
\end{equation}
Squaring the equation and remembering that 
\begin{equation}
Z=C_{\mathrm{crit}}=\sqrt{\tau^{2}+\pi^{2}\sin^{2}\theta_{\mathrm{crit}}^{(d)}}\label{eq:Z_via_tau}
\end{equation}
 (where the sign of the square root is dictated by the fact that $C\geq0$),
we obtain
\begin{equation}
\frac{\tau^{2}}{\sinh^{2}\tau}=\pi^{2}\sin^{2}\theta_{\mathrm{crit}}^{(d)}\iff\frac{\tau}{\sinh\tau}=\pm\pi\sin\theta_{\mathrm{crit}}^{(d)}.\label{eq:tau/sinh=00003D...}
\end{equation}

Recall that only $\tau$ being real or imaginary are the cases of
interest for us. For real $\tau$, $\tau/\sinh\tau\geq0$, dictating
the choice of ``$+$'' in Eq.~(\ref{eq:tau/sinh=00003D...}). Using
this and Eq.~(\ref{eq:Z_via_tau}), we rewrite Eq.~(\ref{eq:cosh+...=00003D0})
as
\begin{multline}
\cosh\tau+\frac{\sqrt{\tau^{2}+\pi^{2}\sin^{2}\theta_{\mathrm{crit}}^{(d)}}}{\pi\sin\theta_{\mathrm{crit}}^{(d)}}\\
=\cosh\tau+\sqrt{\sinh^{2}\tau+1}\\
=2\cosh\tau\geq2>0,
\end{multline}
implying that there are no solutions with real $\tau$.

We thus look for solutions with $\tau=ib$, $b\in\mathbb{R}$. Since
$\tau$ and $-\tau$ are equivalent in Eq.~(\ref{eq:cosh+...=00003D0}),
we choose $b\geq0$ without loss of generality. Then Eq.~(\ref{eq:tau/sinh=00003D...})
becomes 
\begin{equation}
\frac{b}{\sin b}=\pm\pi\sin\theta_{\mathrm{crit}}^{(d)}\iff\frac{\sin b}{b}=\pm\frac{1}{\pi\sin\theta_{\mathrm{crit}}^{(d)}}.\label{eq:sin(b)/b=00003D...}
\end{equation}
The r.h.s. of Eq.~(\ref{eq:sin(b)/b=00003D...}) is either $\geq1/\pi$
or $\leq-1/\pi$. One sees from Fig.~\ref{fig:imaginary-part-crossing_PS}(b)
that $\abs{\sin b/b}\geq1/\pi$ only for $b\leq b_{c}<\pi$. In this
range of $b$, $\sin b/b>0$; therefore, one has to choose ``$+$''
in Eq.~(\ref{eq:sin(b)/b=00003D...}). Not any solution of Eq.~(\ref{eq:sin(b)/b=00003D...})
with ``$+$'' is a solution of Eq.~(\ref{eq:cosh+...=00003D0}).
Indeed,
\begin{multline}
\cosh\tau+Z\frac{\sinh\tau}{\tau}\\
=\cos b+\frac{\sin b}{b}\sqrt{\pi^{2}\sin^{2}\theta_{\mathrm{crit}}^{(d)}-b^{2}}\\
=\cos b+\sqrt{1-\frac{b^{2}}{\pi^{2}\sin^{2}\theta_{\mathrm{crit}}^{(d)}}}\\
=\cos b+\sqrt{1-\sin^{2}b}\\
=\cos b+\abs{\cos b}.
\end{multline}
Therefore, only the $b$ yielding $\cos b<0$ will be actual solutions
of Eq.~(\ref{eq:cosh+...=00003D0}). One can see from Fig.~\ref{fig:imaginary-part-crossing_PS}(b)
that these are $b\in[\pi/2;b_{c}]$.

At $b=b_{c}$, $\sin b/b=1/\pi$ implying $\theta_{\mathrm{crit}}^{(d)}=\pi/2$,
while $b=\pi/2$ implies $\sin\theta_{\mathrm{crit}}^{(d)}=1/2$.
Taking into account Eq.~(\ref{eq:crit_A_theta_postselected}), one
sees that for $A\geq0$ only $\theta_{\mathrm{crit}}^{(d=+1)}\in[\pi/2;5\pi/6]$
are allowed (which implies, again through Eq.~(\ref{eq:crit_A_theta_postselected}),
that only $A_{\mathrm{crit}}\leq A_{0}=\pi\sqrt{3}/2$ are possible).
Now one can construct the critical line. For each $\theta_{\mathrm{crit}}^{(d=+1)}$
in this range, $A_{\mathrm{crit}}$ is found from Eq.~(\ref{eq:crit_A_theta_postselected});
at the same time one solves Eq.~(\ref{eq:sin(b)/b=00003D...}) with
``$+$'' numerically to find $b$, which then yields $C_{\mathrm{crit}}=\sqrt{\pi^{2}\sin^{2}\theta_{\mathrm{crit}}^{(d)}-b^{2}}$.
The resulting critical line is shown in Fig.~\ref{fig:critical_line}.

We note that the arguments presented above rely on the initial assumption
that $P_{\{r_{k}=0\}}^{(d)}$ can only turn to $0$ when Eq.~(\ref{eq:crit_A_theta_postselected})
holds. Abandoning this assumption, there might be, in principle, additional
critical sets $(C_{\mathrm{crit}},A_{\mathrm{crit}},\theta_{\mathrm{crit}}^{(d)})$
that are not included in these considerations. Our numerical investigation,
though, showed no evidence of such points.

\section{\label{sec:appendix_averaged_phase_detection}Details on the averaged
phase detection scheme}

The averaged phase detection setup shown in Fig.~\ref{fig:interferometers}(b)
involves detectors interacting with two arms of the interferometer
via different Hamiltonians. Denoting the upper/lower arm as $a=\pm1$,
we write
\begin{multline}
H_{\mathrm{int}}^{(a)}=-\frac{\lambda(t)}{2}\left(1-a(\mathbf{n}^{(s)}\cdot\bm{\sigma}^{(s)})\right)\\
\times(n_{y}^{(D)}\sigma_{y}^{(D)}+an_{x}^{(D)}\sigma_{x}^{(D)}+an_{z}^{(D)}\sigma_{z}^{(D)}).
\end{multline}
For the upper arm, $a=+1$, this reduces to Eq.~(\ref{eq:Hsd_particular}).
For the lower arm, $a=-1$, which leads to two modifications: the
signs of $n_{x,z}^{(D)}$ are changed and $-\mathbf{n}^{(s)}\cdot\bm{\sigma}^{(s)}$
is measured instead of $\mathbf{n}^{(s)}\cdot\bm{\sigma}^{(s)}$.
This results in a different detector back-action in the lower arm,
given by

\begin{equation}
\tilde{\mathcal{M}}^{(r)}=R^{-1}(\mathrm{\bm{n}}^{(s)})\sigma_{x}^{(s)}\tilde{M}^{(r)}\sigma_{x}^{(s)}R(\mathrm{\bm{n}}^{(s)})\label{eq:our_Kraus_definition-1}
\end{equation}
 with
\begin{eqnarray}
\tilde{M}^{(0)} & = & \begin{pmatrix}1 & 0\\
0 & \cos g-i\sin g\cos\theta^{(D)}
\end{pmatrix}=M^{(0)\dagger},\label{eq:back_action_0_tilde}\\
\tilde{M}^{(1)} & = & \begin{pmatrix}0 & 0\\
0 & -i\sin g\sin\theta^{(D)}e^{-i\varphi^{(D)}}
\end{pmatrix}=M^{(1)\dagger},\label{eq:back_action_1_tilde}
\end{eqnarray}
and the rotation matrix $R(\mathrm{\bm{n}}^{(s)})$ defined in Eq.~(\ref{eq:Hsd_particular}).
The same applies to the last postselected projective measurement,
which is implemented when $g=\theta^{(D)}=-\varphi^{(D)}=\pi/2$.

The particle state just before passing the last beam splitter is 
\begin{multline}
\ket{\Psi}=\frac{1}{\sqrt{2}}\sum_{\{r_{k}\}}\delta_{r_{N+1},0}\ket{\psi_{0}}\prod_{k=1}^{N+1}\ket{r_{k}}_{D_{k}}\\
\times\Biggl[\bra{\psi_{0}}\mathcal{M}_{N}^{(r_{N})}...\mathcal{M}_{1}^{(r_{1})}\ket{\psi_{0}}\ket{+1}_{a}\\
+\bra{\psi_{0}}\sigma_{x}^{(s)}({\bf n}_{0})\tilde{\mathcal{M}}_{N}^{(r_{N})}...\tilde{\mathcal{M}}_{1}^{(r_{1})}\sigma_{x}^{(s)}({\bf n}_{0})\ket{\psi_{0}}\ket{-1}_{a}\Biggr],
\end{multline}
where
\begin{equation}
\sigma_{x}^{(s)}({\bf n}_{0})=R^{-1}({\bf n}_{0})\sigma_{x}^{(s)}R({\bf n}_{0})
\end{equation}
is the ``FLIP'' operator applied twice in the lower arm.

Below we prove that 
\begin{multline}
\bra{\psi_{0}}\sigma_{x}^{(s)}({\bf n}_{0})\tilde{\mathcal{M}}_{N}^{(r_{N})}...\tilde{\mathcal{M}}_{1}^{(r_{1})}\sigma_{x}^{(s)}({\bf n}_{0})\ket{\psi_{0}}\\
=\bra{\psi_{0}}\mathcal{M}_{N}^{(r_{N})}...\mathcal{M}_{1}^{(r_{1})}\ket{\psi_{0}}^{*}.\label{eq:two_arm_ME_relation}
\end{multline}
Using this identity, Eq.~(\ref{eq:averaged_intensity}) for the intensities
at the interferometer's exits immediately follows.

\textbf{\emph{The proof of Eq.~(\ref{eq:two_arm_ME_relation})}}

Note several identities. First,
\begin{multline}
\bra{\psi_{0}}\mathcal{M}_{N}^{(r_{N})}...\mathcal{M}_{1}^{(r_{1})}\ket{\psi_{0}}\\
=\begin{pmatrix}1 & 0\end{pmatrix}\delta RM^{(r_{N})}...\delta RM^{(r_{1})}\delta R\begin{pmatrix}1\\
0
\end{pmatrix},\label{eq:PSME_using_deltaR}
\end{multline}
which follows from definitions of $\ket{\psi_{0}}$ in Eq.~(\ref{eq:psi_0}),
$\mathcal{M}_{k}^{(r_{k})}$ in Eq.~(\ref{eq:our_Kraus_definition}),
and $\delta R$ in Eq.~(\ref{eq:deltaR}). Second,
\begin{equation}
\sigma_{x}^{(s)}\delta R\sigma_{x}^{(s)}=\exp\left(-\frac{2\pi id}{N+1}\right)\sigma_{z}^{(s)}\delta R^{\dagger}\sigma_{z}^{(s)}.\label{eq:deltaR_conjugation}
\end{equation}
Finally,
\begin{multline}
\begin{pmatrix}1 & 0\end{pmatrix}\delta RM^{(r_{N})}...\delta RM^{(r_{1})}\delta R\begin{pmatrix}1\\
0
\end{pmatrix}\\
=\left[\begin{pmatrix}1 & 0\end{pmatrix}\delta RM^{(r_{1})}...M^{(r_{N})}\delta R\begin{pmatrix}1\\
0
\end{pmatrix}\right]^{T}.\label{eq:PSME_transposed}
\end{multline}

Using Eq.~(\ref{eq:deltaR_conjugation}), we show
\begin{multline}
\bra{\psi_{0}}\sigma_{x}^{(s)}({\bf n}_{0})\tilde{\mathcal{M}}_{N}^{(r_{N})}...\tilde{\mathcal{M}}_{1}^{(r_{1})}\sigma_{x}^{(s)}({\bf n}_{0})\ket{\psi_{0}}\\
=\begin{pmatrix}1 & 0\end{pmatrix}\sigma_{x}^{(s)}\delta R\sigma_{x}^{(s)}\tilde{M}^{(r_{N})}...\tilde{M}^{(r_{1})}\sigma_{x}^{(s)}\delta R\sigma_{x}^{(s)}\begin{pmatrix}1\\
0
\end{pmatrix}\\
=\begin{pmatrix}1 & 0\end{pmatrix}\sigma_{z}^{(s)}\delta R^{\dagger}\sigma_{z}^{(s)}\tilde{M}^{(r_{N})}...\tilde{M}^{(r_{1})}\sigma_{z}^{(s)}\delta R^{\dagger}\sigma_{z}^{(s)}\begin{pmatrix}1\\
0
\end{pmatrix}.
\end{multline}
Since $\sigma_{z}^{(s)}\tilde{M}^{(r)}\sigma_{z}^{(s)}=\tilde{M}^{(r)}$
and $\begin{pmatrix}1 & 0\end{pmatrix}\sigma_{z}^{(s)}=\begin{pmatrix}1 & 0\end{pmatrix}$,
\begin{multline}
\begin{pmatrix}1 & 0\end{pmatrix}\sigma_{z}^{(s)}\delta R^{\dagger}\sigma_{z}^{(s)}\tilde{M}^{(r_{N})}...\tilde{M}^{(r_{1})}\sigma_{z}^{(s)}\delta R^{\dagger}\sigma_{z}^{(s)}\begin{pmatrix}1\\
0
\end{pmatrix}\\
=\begin{pmatrix}1 & 0\end{pmatrix}\delta R^{\dagger}\tilde{M}^{(r_{N})}...\delta R^{\dagger}\tilde{M}^{(r_{1})}\delta R^{\dagger}\begin{pmatrix}1\\
0
\end{pmatrix}.
\end{multline}
Using $\tilde{M}^{(r)}=M^{(r)\dagger}$, together with Eqs.~(\ref{eq:PSME_using_deltaR},
\ref{eq:PSME_transposed}), we show
\begin{multline}
\begin{pmatrix}1 & 0\end{pmatrix}\delta R^{\dagger}\tilde{M}^{(r_{N})}...\delta R^{\dagger}\tilde{M}^{(r_{1})}\delta R^{\dagger}\begin{pmatrix}1\\
0
\end{pmatrix}\\
=\left[\begin{pmatrix}1 & 0\end{pmatrix}\delta RM^{(r_{1})}...M^{(r_{N})}\delta R\begin{pmatrix}1\\
0
\end{pmatrix}\right]^{\dagger}\\
=\left[\begin{pmatrix}1 & 0\end{pmatrix}\delta RM^{(r_{N})}...M^{(r_{1})}\delta R\begin{pmatrix}1\\
0
\end{pmatrix}\right]^{*}\\
=\bra{\psi_{0}}\mathcal{M}_{N}^{(r_{N})}...\mathcal{M}_{1}^{(r_{1})}\ket{\psi_{0}}^{*},
\end{multline}
which proves Eq.~(\ref{eq:two_arm_ME_relation}).

\bibliography{bibliography}

\end{document}